\pdfoutput=1
\documentclass[aps,prx,%
    twocolumn,superscriptaddress,longbibliography,floatfix]{revtex4-2}

\usepackage{color}
\usepackage{braket}
\usepackage{amsmath}
\usepackage{esint}
\usepackage{amssymb}
\usepackage{physics}
\usepackage{csquotes}
\usepackage{bm}
\usepackage{graphicx}
\usepackage{epstopdf}
\usepackage{tikz}
\usetikzlibrary{calc}
\usepackage{soul}
\usepackage{xcolor}
\usepackage{bookmark}
\usepackage{array}
\usepackage{slashed}
\usepackage{bbm}
\usepackage{microtype}
\usepackage{upgreek}

\newcommand{\Berry}{{F}}
\newcommand{\Chern}{{\mathcal C}}

\newcommand{\flux}{\text{flux}}
\newcommand{\ext}{\text{ext}}
\newcommand{\drive}{\text{drive}}
\newcommand{\labo}{\text{labo}}
\newcommand{\rot}{\text{rot}}

\begin{document}

\title{
Topological power pumping in quantum circuits
}

\author{J. Luneau}
\email{jacquelin.luneau@ens-lyon.fr}
\affiliation{Univ Lyon, ENS de Lyon, CNRS, Laboratoire de Physique, F-69342 Lyon, France}
\author{C. Dutreix}
\affiliation{Univ. Bordeaux, CNRS, LOMA, UMR 5798, F-33405 Talence, France}
\author{Q. Ficheux}
\affiliation{Department of Physics, ETH Z{\"u}rich, CH-8093 Z{\"u}rich, Switzerland}
\author{P. Delplace}
\affiliation{Univ Lyon, ENS de Lyon, CNRS, Laboratoire de Physique, F-69342 Lyon, France}
\author{B. Dou{\c c}ot}
\affiliation{Laboratoire de Physique Th{\'e}orique et Hautes Energies, Sorbonne Universit{\'e} and CNRS UMR 7589, 4 place Jussieu, 75252 Paris Cedex 05, France}
\author{B. Huard}
\affiliation{Univ Lyon, ENS de Lyon, CNRS, Laboratoire de Physique, F-69342 Lyon, France}
\author{D. Carpentier}
\affiliation{Univ Lyon, ENS de Lyon, CNRS, Laboratoire de Physique, F-69342 Lyon, France}

\date{\today}

\begin{abstract}
In this article, we develop a description of topological pumps as slow classical dynamical variables coupled by a quantum system.  
We discuss the cases of quantum Hall pumps, Thouless pumps, and the more recent Floquet pumps based frequency converters. 
This last case corresponds to a quantum topological coupling between classical modes described by action-angle variables on which we focus. 
We propose a realization of such a topological coupler 
based on a superconducting qutrit suitably driven by three modulated drives. A detailed experimental protocol allowing to measure the quantized topological power transfer between the different modes of a superconducting circuit is discussed. 
\end{abstract}

\maketitle

\section{Introduction}

Topological properties of matter have always been discussed in relation with  topological pumping. 
In an enlightening {\it gedankenexperiment}, B. Laughlin related the topological nature of the 
transverse conductivity of the dimension $D=2$ quantum Hall state to a transfer of charge between 
two edges in a Corbino geometry \cite{laughlin1981quantized}. 
A modern interpretation views the Hall sample as effectively wrapped on a cylinder, realizing 
a $D=1$ quantum Hall charge pump as the enclosed magnetic flux is smoothly increased \cite{kane2013topological,simon2000proposal}. 
 
This quantized adiabatic pumping of a time-dependent quantum system was soon generalized 
beyond the quantum Hall effect to a driven one-dimensional crystal by D. Thouless \cite{thouless1983quantization}.
In a Thouless pump, exemplified by the Rice-Mele model \cite{rice1982elementary}, 
the time-varying flux of the quantum Hall effect is replaced
 by a time-periodic phase $\phi(t)=\omega t$ which accounts generically for the external drive. 
The corresponding dynamics is periodic both in the linear spatial dimension of the crystal, but also in time.
In contrast with previous implementations of geometrical pumps, such topological pumps are characterized by a Chern number and were only recently realized 
 using cold atoms lattices~\cite{lohse2016thouless,nakajima2016topological,lohse2018topological}, 
optical waveguides~\cite{ke2016topological,cerjan2020thouless,jurgensen2021quantized}, magnetically coupled mechanical resonators~\cite{grinberg2020robust} or stiffness-modulated elastic plates~\cite{riva2020edge}. 

An extension of this pumping was unveiled recently in quantum systems of effective spatial dimension $D=0$, 
but with two temporal dimensions corresponding to a drive at two different frequencies $\omega_1$ and $\omega_2$. 
The initial theoretical proposal arose  in Cooper pair pumps realized in Josephson junction driven by 
two independent superconductors' phase differences \cite{riwar2016multi,eriksson2017topological,erdman2019fast,fatemi2021weyl,herrig2020minimal,peyruchat2021transconductance} and later on 
as a frequency converter in a driven  two level system \cite{martin2017topological}. 
This led to the realization of such a frequency converter using a single nitrogen-vacancy center in 
diamond \cite{boyers2020exploring} or with a superconducting qubit \cite{malz2021topological}, 
although in both cases measuring the topological transfer was beyond experimental reach.

In all existing realizations, the topological pumping is probed indirectly, as a property of the 
time-dependent quantum system. 
The dynamics of these quantum systems is assumed to be adiabatic, {\it i.e.} restricted to an eigensubspace well-separated in energy from other quantum eigenstates. 
Pumping manifests itself as 
an anomalous velocity in real space 
for the Quantum Hall and Thouless pumps, or an anomalous velocity in the harmonic Floquet space for the frequency converter.
This anomalous velocity was initially identified in the case of the quantum Hall effect in a crystal~\cite{Niu:1985,Aoki:1986,Niu:1987}, and originates from a Berry curvature whose average value defines a Chern number characterizing the topological nature of the pump. 
It was recently measured in the cold atom~\cite{lohse2016thouless,nakajima2016topological,lohse2018topological} 
and optical waveguides~\cite{cerjan2020thouless} realizations of a Thouless pump.
  
Alternatively to these bulk measurements, the topology of the $D=1$ pumps can be inferred from the occurrence of states at the boundary of the chain during a  period of the drive. 
A direct probe of evolution of these edge states allows to characterize the pumping \cite{grinberg2020robust}. 
Indeed, these topological edge states manifest themselves in the scattering description of the quantum pump~\cite{buttiker1994current,brouwer1998scattering,shutenko2000mesoscopic,avron2000geometry,avron2004transport}: 
they lead to resonances of the reflection matrix $R$ during one cycle of the driven quantum chain, whose associated $\pi$ phase shift lead to the winding of the phase of $\mathrm{det} R$ \cite{simon2000proposal}. 
Such a theoretical approach was recently extended to generalized pumps obtained by wrapping  $D=2$ topological insulators around a cylinder \cite{meidan2011topological,fulga2012scattering}. 

Although previous descriptions of pumping focused on the driven quantum system, they call for a key generalization that embeds the coupled degrees of freedom of the environment if one wants to model the observable topological transfer itself.
In this article, we present such a global framework that allows to characterize the actual pumping and propose  
a realistic experiment that directly measures this topological pumping. 
In practice, a classical treatment of the degrees of freedom of the environment is sufficient to observe a topological transfer and we restrict our attention to that limit.
While a similar treatment dates back to the Born-Oppenheimer approximation \cite{messiah1962quantum,berry1989quantum,berry1993chaotic}, here we focus on classical modes or action-angle variables which describe the periodic drive of a quantum system. 
The interaction between these slow classical degrees of freedom and the gapped quantum system, when treated within the adiabatic approximation, effectively reduces on average
to a topological coupling leading to a quantized pumping between the classical environments.

To illustrate the virtues of this formalism, we propose a realization of a Chern mode coupler 
enabling to measure the topologically protected power transferred  between electromagnetic modes. 
The proposed setup consists in using a qutrit, realized from the first three levels of a highly anharmonic superconducting circuit, a  fluxonium,  driven at multiple microwave frequencies. 
The driven fluxonium realizes a time-dependent version of a Haldane model on a Lieb lattice. 
The corresponding phase diagram can directly be revealed by measuring the power transfer between the microwave modes. The topological pumping leads to a topological redistribution of energy between the 
three microwave   modes.

\section{Slow classical modes coupled to a fast quantum state}
\label{sec:Adiabatic}

\subsection{Dynamics of classical degrees of freedom adiabatically coupled to a quantum system}

\begin{figure}[ht]
\includegraphics[trim = 0mm 0mm 0mm 0mm,clip,width=9cm]{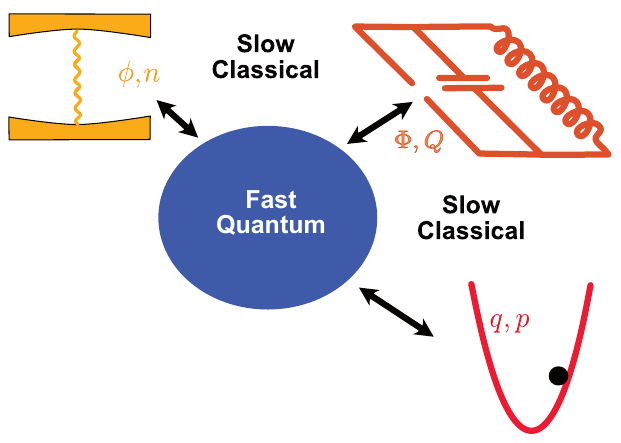}
\caption{
We consider a general gapped and fast system (in blue) coupled to slow variables of the environment (in red). 
The energy separation of the fast quantum system is assumed to be much larger than that of the slow degrees of freedom, allowing a description of the latter by classical pairs of conjugate variables. 
While of different nature, we denote generically these variables by $q_\alpha,p_\alpha$. 
The dynamics of each pair follows from 
a classical Hamiltonian $\mathcal H_\alpha$. The quantum system couples instantaneously to a single 
 variable $q_\alpha$ of each degree of freedom of the classical environment.
}
\label{fig:GeneralSetup}
\end{figure}

We provide a general model of coupled fast and slow degrees of freedom by resorting to a mixed classical - quantum
description, see Fig.~\ref{fig:GeneralSetup}, 
similar in spirit to that used by M.V. Berry and J. Robbins \cite{berry1993chaotic} and by Q. Zhang {\it et al.} \cite{zhang2006general}. 
The fast degrees of freedom are generically described by a quantum 
Hamiltonian~$H$.
The slow degrees of freedom are associated to an energy separation (much) smaller than that of 
the fast system, allowing for a classical description by pairs of conjugated variables~$q_\alpha,p_\alpha$, $\alpha = 1,...N$ satisfying the Poisson bracket relations 
$\{ q_\alpha , p_\beta \} = \delta_{\alpha \beta}$ \cite{Landau:1976}. 
We consider the common case where the quantum system couples to only one of the variable, $q_\alpha$, of each pair of conjugate variables, which includes in particular the case of a driven quantum system.

\subsubsection{Hamilton equations of motion}
The dynamics of each pair of classical variables, prior to the coupling to the quantum system, is assumed to be slow and described by classical dynamics deduced from  a classical Hamiltonian 
${\mathcal H}_\alpha (q_\alpha , p_\alpha)$ following 
\begin{align}
        \dot{q}_\alpha^{(0)} & = 
     \frac{\partial {\mathcal H}_\alpha}{\partial p_\alpha}     
    \label{eq:Hamilton}
    ~~~~~\text{and}~~~~~
    \dot{p}_\alpha^{(0)} =  
    - \frac{\partial {\mathcal H}_\alpha}{\partial q_\alpha} .
\end{align}

The quantum system is described by a Hamiltonian~$H(\{ q_\alpha\})$ parametrized by the 
states of the classical variables~$q_\alpha$. 
We focus on quantum systems that remain gaped during the evolution of the classical modes. 
This allows to approximate at shorter times the dynamics of the quantum system by an 
adiabatic evolution, driven by the slow dynamics of the classical modes. 
More precisely, we consider a quantum system prepared at time $t=0$ 
into one of the eigenstates denoted $\ket{\psi_{\nu}(t=0)}$ 
of the $t=0$ Hamiltonian $H(\{ q_\alpha (t=0)\})$. 
This amounts to assume an initial correlation between the state $\{ q_\alpha (t=0)\}$ 
of the classical environment and the quantum system. 
Through the coupling to $q_\alpha(t)$, 
the quantum system will slowly evolve in time on a timescale dictated by the slow environment. 
We denote its instantaneous state $\ket{\Psi (t)}$ and assume that it remains approximately within the same eigensubspace of the Hamiltonian, which is the essence of the \emph{adiabatic approximation}.
In return, the coupling of the classical degrees of freedom to the quantum system also perturbs their dynamics and results in an effective coupling between different pairs of slow degrees of freedom which is the focus of this paper. 
 
The modified equations of motion for the slow variables are 
\begin{align}
\dot{q}_\alpha & = 
    \dot{q}_\alpha^{(0)}
    \label{eq:Hamilton1}
\\
\dot{p}_\alpha &=  
    \dot{p}_\alpha^{(0)}
    - \bra{\Psi (t)} \frac{\partial H}{\partial q_\alpha} \ket{\Psi (t)}.
    \label{eq:Hamilton2}
\end{align}
The second term in \eqref{eq:Hamilton2} was discussed as a geometric force when $q_\alpha$ is a position in \cite{berry1993chaotic}.
This decomposition of dynamics into  slow and fast degrees of freedom, in the spirit of the Born-Oppenheimer decomposition~\cite{born1954dynamical, messiah1962quantum, mead1979determination}, 
reduces to the standard semi-classical analysis for the specific case of the motion of electrons in insulators~\cite{xiao2010berry}. 

Let us stress that the above description ensures the conservation of the total energy of both quantum and classical degrees of freedom. 
Indeed, the Schr\"odinger equation in finite dimension 
corresponds to classical Hamiltonian equations associated to an Hamiltonian function~$E(\Psi)=\bra{\Psi}H\ket{\Psi}$ and to a natural Poisson bracket structure on the Hilbert space~\cite{heslot1985quantum}.
Therefore the total phase space
also inherits a Poisson bracket structure and the above equations of motion from the 
Hamiltonian function~$\mathcal{H}(\{q_\alpha\},\{p_\alpha \},\Psi)=\sum_\alpha \mathcal{H}_\alpha(q_\alpha,p_\alpha)+\bra{\Psi}H(\{q_\alpha\})\ket{\Psi}$, 
thus abiding by the conservation of the total energy.

To evaluate the matrix elements in~\eqref{eq:Hamilton2}, we now describe the adiabatic evolution of the quantum state~$\ket{\Psi(t)}$.
Due to the slow evolution of the Hamiltonian~$H(\{q_\alpha(t)\})$, 
the state $\ket{\Psi(t)}$ of the quantum system does not identify with the instantaneous eigenstate 
$\ket{\psi_{\nu}(t)}$ defined by 
$H(\{q_\alpha(t)\})\ket{\psi_{\nu}(t)}=
E_{\nu}(\{q_\alpha(t)\})\ket{\psi_{\nu}(t)}$.
The corresponding correction to the dynamics of the slow classical variables~$p_\alpha$ 
in Eq.~\eqref{eq:Hamilton2} is now expressed as  
\begin{align}
\dot{p}_\alpha & = 
    \dot{p}_\alpha^{(0)} 
    - \frac{\partial E_\nu}{\partial q_\alpha}
    + \hbar \sum_{\beta\neq\alpha}  \dot{q}^{(0)}_{\beta} \Berry^{(\nu)}_{q_\alpha q_\beta}, 
\label{eq:MostGeneralMotion2}
\end{align}
see Appendix~\ref{annexe:AdiabEvol} for the detailed derivation.
The first correction relates to the energy variation of the quantum state and follows from the standard classical - quantum coupling. 
The last correction is more exotic: it manifests an effective coupling between 
the different slow variables $q_\alpha ,p_\alpha$. The strength of this transverse coupling  
depends on the geometry of the eigenstates $\ket{\psi_{\nu}}$ of the quantum system 
through the components $\Berry_{q_\alpha q_\beta}^{(\nu)}$ of the  two-form Berry curvature 
which are defined as 
\begin{multline}
    \Berry_{q_\alpha q_\beta}^{(\nu)}= \\
    i\sum_{\mu\neq \nu}
    \frac{
    \bra{\psi_{\nu}}  \partial_{q_\alpha}H\ket{\psi_{\mu}}
    \bra{\psi_{\mu}} \partial_{q_\beta}H\ket{\psi_{\nu}} 
    }{(E_{\nu}-E_{\mu})^2} - (\alpha \leftrightarrow \beta).
\label{eq:curvature}
\end{multline}

\subsubsection{Condition of adiabaticity}
\label{sec:conditionAdiab}

The above adiabatic approximation in one eigensubspace~$\ket{\psi_\nu}$ 
holds as long as the transitions to a different eigensubspace~$\ket{\psi_\mu}$ can be neglected.
These non-adiabatic transitions can be described as Landau-Zener transitions
~\cite{zener1932non, landau1932theory}, 
with a probability of transition from one eigenstate to another behaving as 
$\exp(-\pi/(4\max_t\epsilon_{\mu\nu}))$
where the time dependent parameter~$\epsilon_{\mu\nu}$ reads 
\begin{equation}
    \epsilon_{\mu\nu} = 
    \hbar\frac{\left| \bra{\psi_\mu}\frac{\dd H}{\dd t}\ket{\psi_\nu}\right|}{(E_\mu-E_\nu)^2}.
\end{equation}
In this formula,  $\tau_{\mu\nu}^\mathrm{col}=|E_\mu-E_\nu|/\left| \bra{\psi_\mu}\frac{\dd H}{\dd t}\ket{\psi_\nu}\right|$ corresponds to the characteristic time of the Landau-Zener collision, 
or equivalently to an energy spread~$\delta E=\hbar/\tau_{\mu\nu}^\mathrm{col}$. 
Transitions occur when this spread is comparable to the gap~$E_\mu-E_\nu$, and the parameter 
$\epsilon_{\mu\nu}$, which controls the validity of the adiabatic approximation,   
identifies with this ratio~$\epsilon_{\mu\nu}=\delta E/|E_\mu-E_\nu|$.
For an alternative derivation of this small parameter~$\epsilon_{\mu\nu}$ within the 
adiabatic expansion~\cite{mostafazadeh1997quantum}, see Appendix~\ref{annexe:AdiabEvol}.

Along a trajectory in phase space, the gap of the quantum system varies. The transition dynamics at each minimum can be described following the above Landau-Zener approach, leading to a series of transitions. 
It is then possible to determine
the characteristic time of validity of the adiabatic approximation~$\tau_\mathrm{adiab}$ by constraining 
 the cumulative transition probability to be {\it e.g.} of order~$0.1$. 
Introducing the mean free time~$\tau_\mathrm{mft}$
separating the evolution in phase space between two minima of the gap, we get
\begin{equation}\label{eq:adiabTime}
    \tau_\mathrm{adiab} \approx  0.1 ~ \tau_\mathrm{mft}~ \exp\left(\frac{\pi}{4\max_t \epsilon_{\mu\nu}}\right) 
\end{equation}
where the maximum of the parameter $\epsilon_{\mu\nu}$ is evaluated along the phase space trajectory during $\tau_\mathrm{mft}$. 
Note that for the above analysis to be consistent, the collision time~$\tau_{\mu\nu}^\mathrm{col}$ must be smaller than the mean free time~$\tau_\mathrm{mft}$. 
Since we focus on aperiodic evolution, we also neglected the effect of relative phases accumulated between the transitions~\cite{shevchenko2010landau}.

\subsection{Geometrical power transfer}

The geometrical coupling in~\eqref{eq:MostGeneralMotion2}
between the different subset of slow variables  $q_\alpha (t),p_\alpha (t)$ is associated with an 
energy transfer between them.  
The change of energy of each classical degree of freedom is: 
\begin{align}
    \frac{\dd\mathcal{E}_\alpha}{\dd t} 
    &= \dot{q}_\alpha  \frac{\partial\mathcal{H}_\alpha}{\partial q_\alpha} 
        + \dot{p}_\alpha  \frac{\partial\mathcal{H}_\alpha}{\partial p_\alpha} 
    \nonumber \\
    &= -\dot{q}_\alpha^{(0)}\frac{\partial E_\nu}{\partial {q_{\alpha}} }  
    + \hbar \sum_{\beta\neq\alpha}  \dot{q}^{(0)}_{\alpha}\dot{q}_\beta^{(0)}  \Berry^{(\nu)}_{q_\alpha q_\beta}.
\label{eq:SimplifiedEnergyTransfer}
\end{align} 
The antisymmetry of the Berry curvature implies the conservation of the total energy.

\subsection{Nature of classical degrees of freedom}

Let us discuss briefly the implications of the previous modification of Hamilton equations for different types of classical slow degrees of freedom coupled to a quantum system. 

\subsubsection{Massive classical particles}

The initial context of the Born-Oppenheimer approximation, at the origin of the adiabatic approximation, was the description of light particles, the electrons, coupled to heavy particles,  the nucleus. 
In this situation, the slow degrees of freedom described classically are those of the massive 
particle: its position $q_\alpha$ and conjugated momentum $p_\alpha$. 
The corresponding Hamiltonian is 
$\mathcal{H}_\alpha = p_\alpha^2/(2M) + V( q_\alpha) $, 
parametrized by the mass $M$ and potential $V(q)$. 
The equations of motion in this case take the form 
\begin{align}
\dot{q}_\alpha & = 
    \frac{p_\alpha}{M},
\label{eq:GeneralMotionMassive1}
\\
\dot{p}_\alpha & = 
    -V'(q_\alpha) 
    - \partial_{q_{\alpha}} E_\nu
    +\hbar\sum_{\beta\neq\alpha}
    \frac{p_\beta}{M}
    \Berry^{(\nu)}_{q_\alpha q_\beta}.
\label{eq:GeneralMotionMassive2}
\end{align} 
Equation~\eqref{eq:GeneralMotionMassive2} describes the associated anomalous geometrical
force~\cite{berry1989quantum,berry1993chaotic}.

\subsubsection{Classical modes}
\label{sec:ClassicalModes}

While the previous adiabatic formalism was initially designed with classical massive degree of freedom in mind, it also applies to the case of slow action angle $\phi_\alpha , n_\alpha$ variables, 
which we will call a \emph{classical mode}. 
In more details, we consider a variable $p_\alpha = \hbar n_\alpha $ where $n_\alpha$ takes only 
integer values 
and its canonical phase  
$q_\alpha = \phi_\alpha$, a $2 \pi$ periodic phase. 
Of particular interest is the situation of a monochromatic mode, corresponding to the 
Hamiltonian 
$\mathcal H_\alpha = \hbar \omega_\alpha n_\alpha = \omega_\alpha p_\alpha $ 
whose linearity in $n_\alpha$ is the distinctive feature compared with the massive case. 
In this situation, the modified Hamilton equations of motion read 
\begin{align}
\dot{\phi}_\alpha = \dot{q}_\alpha & =  
    \omega_\alpha
    , 
\label{eq:GeneralMotionMode1}
\\
\hbar \dot{n}_\alpha =\dot{p}_\alpha & = 
    - \frac{\partial E_\nu}{\partial \phi_{\alpha} } 
    + \hbar \sum_{\beta\neq\alpha}  \omega_{\beta} \Berry^{(\nu)}_{\phi_\alpha \phi_\beta}
    . 
\label{eq:GeneralMotionMode2}
\end{align}
In the case of classical modes, 
the Eq.~\eqref{eq:GeneralMotionMode2} describes the filling rate of mode $\alpha$. 
The energy transferred between the different modes corresponds to 
\begin{align}
    \frac{\dd\mathcal{E}_\alpha}{\dd t} =& 
     -\omega_\alpha \frac{\partial E_{\nu}}{\partial \phi_\alpha} 
    + \hbar \sum_{\beta\neq\alpha} \omega_\alpha\omega_\beta \Berry^{(\nu)}_{\phi_\alpha \phi_\beta} . 
\label{eq:EnergyTransferModes}
\end{align}

\section{Topological pumping}

\subsection{Topological versus geometrical couplings}

In the above section, we have shown how a geometrical quantity, the Berry curvature, encodes the strength of the effective coupling mediated by a quantum system between classical variables. 
Of particular interest is the case of two classical variables, {\it e.g.} $q_1$ and $q_2$ with a compact configuration space
denoted $[0,2\pi]^2$. 
In this situation, introducing the integer $\Chern_{12}^{(\nu)} $ Chern number, 
the quantity 
\begin{equation}
    \int_{[0,2\pi]^2} \dd q_1 \dd q_2 ~ \Berry^{(\nu)}_{q_1 q_2}
    =2\pi ~ \Chern_{12}^{(\nu)} 
\label{eq:ChernNumber}
\end{equation} 
is a topological quantity quantized in units of $2\pi$, 
{\it i.e.} it is insensitive of perturbations of the quantum Hamiltonian~$H$ provided the gap between~$E_\nu $ and other states does not close.
 
Noting that \eqref{eq:ChernNumber} is nothing but the averaged Berry curvature over the configuration space, the case of a Hamiltonian $H$ characterized by a topological Chern number corresponds to a quantized averaged coupling in 
Eq.~\eqref{eq:MostGeneralMotion2}. 
When the Berry curvature fluctuates in the configuration space around its topological average, 
 a quantized coupling between the classical variables is recovered when averaging over initial position in the configuration space, which is hardly practical. 
Instead, we can resort to a time average: if the evolution of classical variables is ergodic, an average over the configuration space can be replaced by an average over a sufficiently long time.  

The above situation of two classical environments topologically coupled through a quantum system correspond to a topological pump. 
Historically, the relation between a topological Chern invariant and a pumping process originates from the Laughlin's description of a charge 
transfer between inner and outer edges of a quantum Hall sample in a Corbino geometry \cite{laughlin1981quantized}. 
Later on, D. Thouless proposed another topological pumping mechanism by periodically modulating a $D=1$ crystal. 
Much more recently,  realizations of a topological pump were proposed 
as a junction between 
superconductors  \cite{riwar2016multi,eriksson2017topological,fatemi2021weyl} 
or by driving a two level system at two different frequencies \cite{nathan2019topological}. 
The topological nature of these processes can be inferred from the dynamics of the 
quantum degrees of freedom: in all cases the corresponding band structure is gapped,
and the eigenmanifold in which the adiabatic dynamics takes place is characterized 
by a Chern number~\eqref{eq:ChernNumber}.
 
Alternatively, such topological pumps can be characterized from a scattering point of view 
by probing the appearance of topological edges  during the evolution of the quantum 
system through phase shifts of the reflection matrix
\cite{simon2000proposal,meidan2011topological,fulga2012scattering}. 
In the following, we develop a ``Born-Oppeiheimer description" of this topological pumping by showing how they can naturally be described with the  
formalism of 
Sec.~\ref{sec:Adiabatic} of adiabatic topological 
coupling between classical slow variables of different nature.

\subsection{$D=2$ Quantum Hall pump}
\begin{figure*}[t]
\includegraphics[trim = 0mm 0mm 0mm 0mm,clip,
            width=160mm]{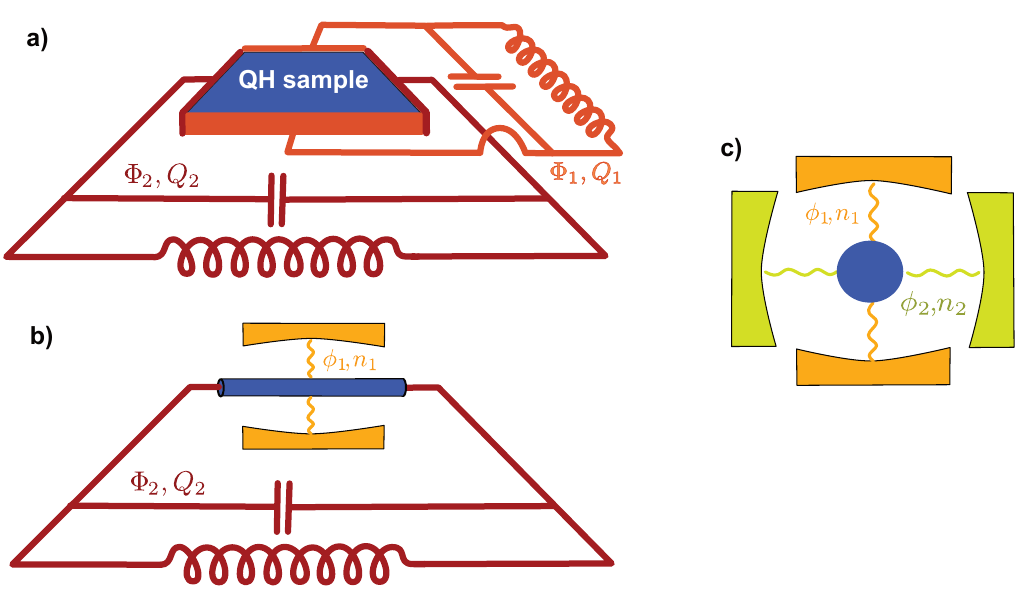}
\caption{%
(a) Schematic quantum Hall circuit where two LC branches, described
by classical conjugate variables $\Phi$ and $Q$, are connected through a quantum Hall 
sample. 
(b) Schematic Thouless pump driven by a phase $\phi_1 = \omega_1 t$ conjugate to a variable $n_1$, and coupled to an  LC branch. Topological pumping gives rise to a current in the $LC$ circuit.
(c) Topological mode coupler, or frequency converter, in which two classical modes 
described by $\phi_1,n_1$ and $\phi_2,n_2$ variables are coupled
topologically through a quantum system. 
}
\label{fig:TopoPump}
\end{figure*}

In an enlightening argument, B. Laughlin related the quantization of the 
transverse conductivity of the quantum Hall state to a transfer of charge between 
 edges in a Corbino geometry as the flux threading the disk is increased by one 
 quantum \cite{laughlin1981quantized}. 
Later on,     
Niu, Thouless and Wu introduced the notion of generalized boundary conditions 
for quantum Hall states \cite{Niu:1985}. 
The quantum Hall topological properties are expressed as the Chern number of the ensemble of many-body groundstates over the closed manifold of 
phase boundary conditions.
These boundary conditions parameters were later related to electromotive forces through loops connecting opposite edges of the sample \cite{avron1983homotopy,avron1987adiabatic}
 effectively generalizing the topology of Laughlin's gedanken experiment 
 to that of a torus and allowing for a dynamical description of the quantum Hall effect over a  classical parameter space \cite{gritsev2012dynamical}.   

Here we consider the classical phases  entering the generalized boundary conditions as dynamical variables. This effectively amounts to realize a quantum Hall topological pump between two $LC$ harmonic circuits. 
Let us consider a quantum Hall sample coupled to two independent electrical circuits 
in the $x$ and $y$ direction, see Fig.~\ref{fig:TopoPump}(a). 
The coupling between each circuit and the quantum Hall sample 
follows from the boundary conditions of Niu {\it et al.}  \cite{Niu:1985} 
on the many-body ground state wave function~$\Psi(x_i,y_i)$: 
\begin{subequations}
\begin{align}
\Psi(x_i + L_{x},y_i) &= e^{i \Phi_{1}  } ~\Psi (x_i,y_i) 
 ,  \\
\Psi(x_i ,y_i+ L_{y}) &= e^{i \Phi_{2} }~\Psi (x_i,y_i) . 
\end{align}
\label{eq:BoundaryConditions}
\end{subequations}
The two phases~$\Phi_1, \Phi_2$ are  
related to the voltage drop~$V_\alpha$ in each directions as  ~$\Phi_\alpha (t)= (e/\hbar) \int^t V_\alpha(t') dt'$ . 
If we model each electric branch associated to this voltage drop as an LC circuit
\cite{devoret1995quantum}, these phases are dynamical classical variables, whose 
canonically conjugate momenta are the (rescaled) accumulated charge in each circuit $Q_\alpha = (\hbar / e) \int^t I_\alpha(t') dt' $, $I_\alpha$ being the current in each circuits.   
The classical Hamiltonian describing each LC circuit is 
\begin{equation}
   \mathcal{H}_\alpha =
   \frac{e^2}{2\hbar^2 C_\alpha}Q_\alpha^2 +
    \frac{\hbar^2}{2e^2 L_\alpha} \Phi_\alpha^2
    ~,
\label{eq:H-LCcircuit}
\end{equation}
where $L_\alpha$ is the inductance and $C_\alpha$ the capacitance of the corresponding circuit~\cite{devoret1995quantum}. 
Hence the dynamics of an LC circuit identifies to that of a massive particle
of position  $q_\alpha=\Phi_\alpha $ and momentum  $p_\alpha = Q_\alpha$, 
in a harmonic potential. 
 
The Hamilton's equations 
(\ref{eq:GeneralMotionMassive1} and \ref{eq:GeneralMotionMassive2}) include a 
correction to the usual relations between flux and current 
\begin{align}
        \dot{Q}_1 &= \frac{\hbar}{e} I_1  
        = -\left( \frac{\hbar}{e}\right)^2  \frac{\Phi_1}{L_1} 
        + \hbar F_{\Phi_1\Phi_2} \frac{e}{\hbar} V_2 
\end{align}
{\it via} the Berry curvature  $F_{\Phi_1\Phi_2}$ of the quantum Hall effect ground states derived by Niu {\it et al.} 
\cite{Niu:1985}. 
This Berry curvature being independent of the external fluxes
and thus constant over the parameter space, it is related to the quantum Hall 
Chern number $\Chern_{12}$ via $F_{\Phi_1\Phi_2} =\Chern_{12}/(2\pi)$. 
Thus the corrected classical equation of motion of the $LC$ circuits reads
\begin{equation}
    I_1 = -\frac{\hbar}{e}  \frac{\Phi_1}{L_1} + \frac{e^2}{h} \Chern_{12} V_2 .  
\end{equation}
The usual case of an ideal ammeter is recovered in the limit $L_1 \to \infty , C_1 \to 0$. 

Note that the energy transferred from one LC circuit to the other, following
\eqref{eq:SimplifiedEnergyTransfer}, is 
\begin{equation}
    \frac{\dd E_1}{\dd t} = \hbar F_{\Phi_1 \Phi_2} \dot{\Phi}_1 \dot{\Phi}_2 
    =  \Chern_{12} \left( \frac{e^2}{h}\right) V_1 V_2 = 
    \delta I_1 V_1 . 
\end{equation}
In the limit of an ideal Hall measurement, where the $L_1C_1$ circuit corresponds to an 
ammeter, we get $V_1=0$, and no energy is transferred between the two circuits.

\subsection{$D=1$ Thouless pump}

Let us now turn to the canonical example of a topological pump, proposed by D. Thouless \cite{thouless1983quantization}, 
which consists in a $D=1$ crystal suitably periodically driven  in time, such as the Rice-Mele model
\cite{rice1982elementary}. 
The single electron dynamics is thus described by a time-dependent Bloch Hamiltonian $H(k,\phi(t))$ periodic both in momentum over the Brillouin zone, and in $2\pi$ periodic phase $\phi(t)=\omega t$. 
 This Hamiltonian is assumed to be gapped at all time, and with energy band $\nu$  eigenstates $\ket{\psi_\nu (k,\phi)}$ 
possessing a finite Chern number $\Chern_{k \phi}^{(\nu)}$ 
over the $2$-torus constituted of the $D=1$ Brillouin zone and periodic phase configuration space of phase $\phi$. 
In such a system,   
topological pumping is usually described as the appearance of a steady current in the bulk of a closed ring,
corresponding to an anomalous geometric velocity for semi-classical states in band $\nu$ 
$\left< \dot{x} \right>  = \langle F_{k \phi}^{(\nu)} \rangle \partial_t \phi = 2\pi~ \Chern_{k \phi}^{(\nu)}~ \omega$
\cite{karplus1954hall,xiao2010berry}. 

We propose an alternative description of such a Thouless pump, by considering an open $D=1$ crystal 
of size $L$ 
connected on both ends to an $LC$ circuit, analogous to the quantum Hall pump, see Fig.~\ref{fig:TopoPump}(b).
The coupling between the charged particles in the crystal and the $LC$ circuit follows from the boundary conditions \eqref{eq:BoundaryConditions} on  the many-body
groundstate wavefunction 
$\Psi(x=0,\phi(t)) = e^{i\Phi_1} \Psi(x=L,\phi(t)) $. 
This amounts to couple the crystal to a pair of classical conjugated variables $Q_1,\Phi_1$ identical to 
those in the quantum Hall pump, with a classical Hamiltonian \eqref{eq:H-LCcircuit} describing their dynamics. 
The periodic driving of the Hamiltonian is now interpreted as the coupling between the charges of the 
crystal and a dynamical classical variable $\phi_2 = \omega t$, conjugated to a variable $n_2$. 
The dynamics $\dot{\phi}_2 = \omega , \dot{n}_2 = 0$, correspond to that of a classical mode introduced in 
Sec.~\ref{sec:ClassicalModes}. 
 
In this representation, the topological coupling between the $LC$ circuit and the classical mode leads to 
modified equation of motion in the $LC$ circuit, manifesting the appearance of a charged current. 
The modified Hamilton equation \eqref{eq:GeneralMotionMassive1} reads 
\begin{equation}
    \dot{Q}_1 = \frac{\hbar}{e} I_1 
        = -\left( \frac{\hbar}{e}\right)^2  \frac{\Phi_1 }{L} 
        + \hbar F_{\Phi_1\phi_2}\omega , 
\end{equation}
which leads to a steady charge current 
$I_1 = e 2\pi F_{\Phi_1\phi_2} / T $ through the driven crystal, 
corresponding to an average number $\Chern_{\Phi_1\phi_2}^{(\nu)}= \Chern_{k \phi}^{(\nu)}$ of 
charged transferred across the chain per period $T$ of the drive.  
This results identifies with the standard steady anomalous velocity 
 in the bulk of the Thouless pump.

\subsection{$D=0$ Power pump} 
\label{sec:genModeCoupl}

In the previous section, we interpreted a time-periodic quantum Hamiltonian as a coupling between a fast quantum system and a  slow classical mode. 
A natural extension consists in considering a single quantum system coupled to the phases $\phi_\alpha$ of 
 an arbitrary number $N$ of classical modes, see Fig.~\ref{fig:TopoPump}(c). 
 The quantum dynamics of such a system can be described within 
 Floquet theory \cite{martin2017topological}. 
 The topological Chern numbers of such a quantum system are defined in Floquet space, and, when non zero, leads to a frequency conversion mechanism.
  
The description of such a \emph{quantum mode coupler} is natural in terms of an effective classical 
dynamics of the modes. 
We assume that the quantum system couples only to the phases of the modes, corresponding to a 
Hamiltonian $H(\phi_1(t), \dots,\phi_N(t) )$. 
In such a case, the equations of motions are given by (\ref{eq:GeneralMotionMode1}, \ref{eq:GeneralMotionMode2}) 
 with a power leaving each mode given by Eq.~\eqref{eq:EnergyTransferModes}. 
In the particular case of two modes of frequency $\omega_1,\omega_2$, 
we recover the result of Martin {\it et al.} for the averaged power 
 between two modes given by 
$\dd \mathcal{E}_1/\dd t = \hbar \omega_1 \omega_2~ \Chern_{\phi_1\phi_2}^{(\nu)} /(2\pi)$~\cite{martin2017topological}. 
 
In the following section, building on this general description of topological pumping 
we propose a realization of such a quantum 
topological coupler between microwaves modes using an artificial $3$ level atom, a qutrit.

\section{Topological pump in quantum circuits : the topological qutrit}
 
\subsection{Case of a qubit}

It is possible to apply the concept of geometrical and topological response of gapped states to individual quantum systems. The seminal work of Martin, Refael, and Halperin~\cite{nathan2019topological} proposed to use a spin-1/2 under two frequency drives to observe the quantized pumping of energy from one drive to the other.
Measuring such a power transfer presents a substantial experimental challenge.
Recently, Malz and Smith~\cite{malz2021topological} used the IBM Quantum Experience
to observe the
inner dynamics of a superconducting qubit state that would correspond to a topological quantum transition. However, their control scheme was mixing the flows of power between the drives -- hindering a direct measurement of the power transfer. Similarly, other experiments on NV centers demonstrate a topological transition in the qubit dynamics but could not explore the quantized pumping of power~\cite{boyers2020exploring}. 

Essentially, the model
proposed in Ref.~\cite{nathan2019topological} consists in engineering the Hamiltonian
\begin{multline}
        H(\phi_1,\phi_2) \propto  \sin(\phi_1)\sigma_X+\sin(\phi_2)\sigma_Y \\
        +\left(M-\cos(\phi_1)-\cos(\phi_2)\right)\sigma_Z,
    \label{eq:HamiltonianMartin}
\end{multline}
where
$\sigma_i$ are the Pauli operators of the qubit, $M$ is a parameter that drives the topological transition and the phases $\phi_1=\omega_1 t$ and $\phi_2=\omega_2 t$ are driven at two incommensurate frequencies. We envision three ways to realize this Hamiltonian.
\begin{itemize}
    \item As in Ref.~\cite{malz2021topological}, it is possible to drive a single superconducting qubit, i.e. a transmon, with a complex amplitude $\left[X(t)+i Y(t)\right]e^{-2i\pi f_q t - 2i\int_0^t Z(\tau)\mathrm{d}\tau }$, where $f_q$ is the qubit frequency in order to implement any driving term of the sort $H(t)=X(t)\sigma_X+Y(t)\sigma_Y+Z(t)\sigma_Z$. However, while it is possible to infer what is the transferred power between frequency components at $\omega_1$ and $\omega_2$ from the measured qubit dynamics, this power flow lacks physical embodiment and cannot be measured using any known apparatus.
    \item Alternatively, the $\sigma_Z$ term in the Hamiltonian (\ref{eq:HamiltonianMartin}) can be achieved by controlling the frequency of the qubit directly, hence physically separating the source of power from the three terms corresponding to each Pauli operator in Eq.~(\ref{eq:HamiltonianMartin}). The frequency of flux tunable qubits can be tuned rapidly using a on-chip flux control. 
    However, measuring the power of the drive used for such a flux-tunable bias has never been achieved to our knowledge and requires involved technical development.
    \item Lastly, the frequency of the qubit can be controlled by exploiting the ac-Stark shift created by a drive far detuned from the qubit transition frequency. The ac-Stark shift is a commonly used method to engineer the spectrum of artificial atoms but the relatively low anharmonicity of transmons imposes a finite bandwidth on the control parameter $Z(t)$.
\end{itemize} 

All these solutions present serious practical limitations either on the achievable Z-control or -- more importantly -- on the ability to measure the quantized power transfer. We propose to circumvent this difficulty by extending the size of the Hilbert space. This new pumping schemes uses a qutrit to create gapped states.

\subsection{Implementation with a superconducting qutrit}

\subsubsection{Principle of the experiment}\label{sec:exp}

\begin{figure*}[t]
\includegraphics[trim = 0mm 0mm 0mm 0mm,clip,width=13cm]{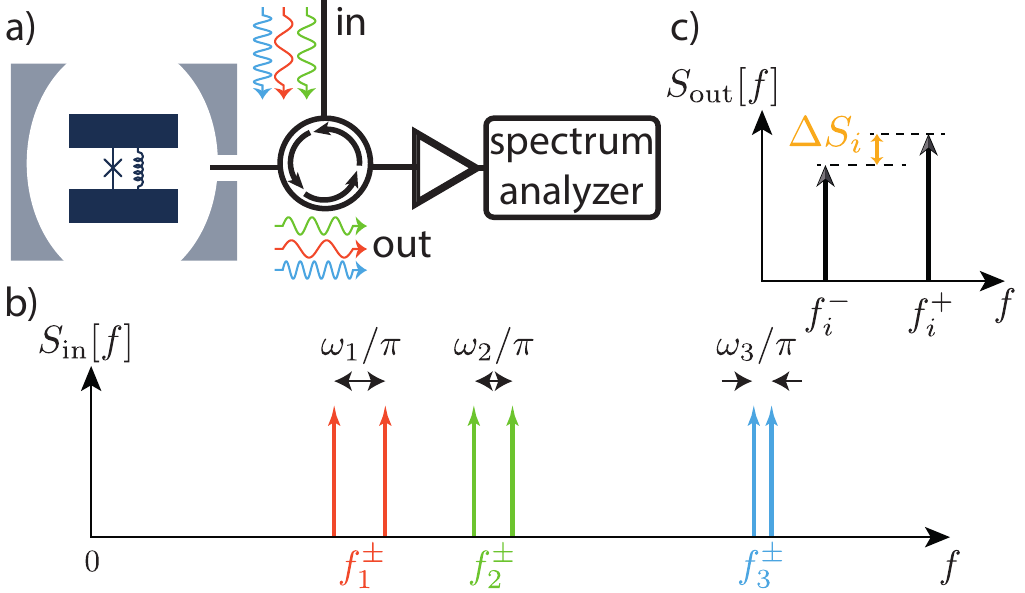}
\caption{
(a) Schematic of the experimental setup. A fluxonium circuit is embedded in a host cavity. The transitions between the first three levels of the fluxonium are driven with a detuning $\delta_i$, an amplitude $\Omega_i$, and a modulation frequency $\omega_i$ (blue, red and green). The power of the outgoing signals are recorded with a power spectrum analyzer that provides the instantaneous photon flux of each frequency mode.
(b) Spectrum of the driving tones. Each fluxonium transition is driven with two side-bands used to implement a topologically protected power transfer.
(c) For decoherence rates smaller than the modulation frequencies $\omega_i$, the power of each sideband can be resolved. The quantized power transferred is expressed as a function of the difference of the spectral power $\Delta S_i$ in the sidebands of each reflected driving tone according to Eq.~(\ref{Eq:tocheckexperimentally}).
}
\label{fig:exp}
\end{figure*}

The experiment we propose consists in driving a superconducting qutrit at several frequencies in order to establish a topologically given power flow between microwave modes at various frequencies. We propose to use a superconducting circuit behaving as a qutrit where every transitions can be addressed individually with a well defined phase. In the following, we denote the transitions $|0\rangle-|1\rangle$ as $1$, $|1\rangle-|2\rangle$ as $2$, and $|0\rangle-|2\rangle$ as $3$ for the sake of simplicity. By modulating the drive amplitude $\Omega_i$ of each transition $i$ at a frequency $\dot{\phi}_i=\omega_i$, it is possible to engineer an effective Hamiltonian in a configuration space defined by the phases $\phi_1, \phi_2, \phi_3$. By enforcing $\phi_3=\phi_1-\phi_2$, the dimension is reduced while still enabling the observation of a topological transition in the power transferred between the various driving tones.
The difference between any two transition frequencies of the qutrit shall be much greater than any drive amplitude $\Omega_i$ and modulation frequency $\omega_i$ in order to enable the direct measurement of the power transfer between any driving modes. Finally, all the above-mentioned timescales should be much smaller than the coherence time of the qutrit transitions.

An example of a superconducting circuit satisfying theses requirements is the fluxonium artificial atom. The fluxonium is a highly anharmonic superconducting circuit whose first three energy levels can be used as a qutrit. The circuit is a loop composed of a Josephson junction shunted by a large inductance, see Fig.~\ref{fig:exp}(a). When the loop is threaded by an external magnetic flux corresponding to almost (but not exactly) half a flux quantum, no selection rule prevents the direct driving of all three transitions while the circuit transitions have been shown to display record long coherence times~\cite{somoroff2021millisecond,zhang2021universal,nguyen2019high}. 
 
The circuit is embedded in a cavity with a single port connected to a transmission line.
The cavity is used as an off-resonant readout mode dispersively coupled to the circuit transitions~\cite{zhu2013circuit}. The cavity also acts as a filter that protects the circuit from direct energy dissipation into the electromagnetic environment of the transmission line, while preserving fast microwave control
through to the direct coupling of the input port to the circuit antenna.

Incoming microwave modes carry the modulated drives [see Fig.~\ref{fig:exp}(b)] to the qutrit and outgoing modes carry the reflected signal before being measured by a power spectrum analyzer. At the input, each drive at frequency~$f_i$ is modulated in amplitude at a frequency~$\omega_i$, resulting in the pairs of sidebands in Fig.~\ref{fig:exp}(b). The geometrical and topological signatures can be observed in the power transfer between these various frequency modes. 
We model the propagating mode in the transmission lines at frequency~$f$ as a classical mode of energy~$hf n_f$ such that the net photon flux is given by the difference between the outgoing and incoming signals at this frequency $(S_\text{out}[f]-S_\text{in}[f])/hf$.
Precisely, one first needs to probe the difference $\Delta S_i$ in power spectral density between two sidebands [see Fig.~\ref{fig:exp}(c)] and convert it in photon flux $\dot{n}_i =\Delta S_i/h f_i$ (see Appendix~\ref{sec:powerToMeasure} for a refined expression).

\subsubsection{Hamiltonian in the rotating frame}\label{sec:HamiltonianDerivation}

The fluxonium Hamiltonian reads~\cite{manucharyan2009fluxonium}
\begin{equation}\label{eq:fluxoniumHam}
    H_\mathrm{fluxonium} = 4 E_C \hat N^2 + \frac{E_L}{2} \hat\varphi^2 - E_J \cos(\hat\varphi-\varphi_\ext),
\end{equation} where $\hat{N}$ is the charge on the capacitor of the circuit, $\hat{\varphi}$ is the phase twist across the inductance, $\varphi_\mathrm{ext}$ is the external magnetic flux threading the loop, and $E_C,~E_L,~E_J$ are 
respectively the charging energy, inductive energy and Josephson energy of the circuit.
The fluxonium is addressed by a microwave drive applied on a capacitance, so that each pump induces a term 
proportional to~$\cos(\phi_i)\cos(\theta_i) \hat N$
in the Hamiltonian, where~$\theta_i(t)=2\pi f_i t + \theta_i^0$ is the phase of each tone and $\phi_i(t)=\omega_i t$ is the phase of their amplitude modulation.
We denote as $\ket{0}, \ket{1}, \ket{2}$ the  first three energy levels of the fluxonium~\eqref{eq:fluxoniumHam}.
The charge operator~$\hat N$ is off-diagonal in this basis~$(\ket{0}, \ket{1}, \ket{2})$. Hence the effect of the drives is purely off-diagonal.
The frequencies of the three drives~$f_1$, $f_2$ and~$f_3$ are constrained to satisfy 
$f_3 = f_1 + f_2$ so that each tone drives a single transition of the fluxonium. This constraint can be enforced in the microwave domain by using mixers to generate a tone at $f_3$ using tones at $f_1$ and $f_2$ or by direct numerical synthesis. 

We move to a rotating frame by applying the unitary diagonal transformation~
$U(t) = \textrm{diag} (1, \exp(-2i\pi f_1 t), \exp(-2i\pi f_3 t) )$.
One can choose the initial phases~$\theta_i^0$ of the tones so that
in the rotating frame and 
using the rotating wave approximation, the dynamics of the qutrit is governed by the Hamiltonian
\begin{multline}
        \tilde{H}(\phi_1, \phi_2, \phi_3)=  \\
    \hbar\left(
    \begin{array}{ccc}
    0 & \Omega_{1}\cos(\phi_1) & -i\Omega_{3}\cos(\phi_3) \\
	\Omega_{1}\cos(\phi_1) & \delta_{1} & \Omega_{2}\cos(\phi_2) \\
	i\Omega_{3}\cos(\phi_3) &\Omega_{2}\cos(\phi_2) & \delta_{3}
	\end{array}
    \right)
    \label{eq:Hphi1phi2phi3}
\end{multline}
which depends on the phases~$\phi_i(t)=\omega_i t$ of the drive amplitude modulations,
where $\delta_{1}=2\pi f_{01}-2\pi f_1$ and $\delta_{3}=2\pi f_{02}-2\pi f_3$ are the frequency detuning between the fluxonium transition frequencies and the drive frequencies (see Appendix~\ref{annexe:RotFrame} for details). 
Note that the above constraint on drive frequencies  sets $\delta_2=\delta_{3}-\delta_{1}$. Besides, in order to simplify the dynamics, we impose an additional constraint on the modulation frequencies, namely $\phi_3=\phi_1-\phi_2$. Therefore the effective Hamiltonian can be described by an Hamiltonian evolution controlled by two phases only $H(\phi_1, \phi_2) \equiv \tilde{H}(\phi_1,\phi_2,\phi_1-\phi_2)$.
The rotating wave approximation is valid if the detunings~$\delta_{i}$ and the drive amplitudes~$\Omega_{i}$ are much lower than the fluxonium transitions frequencies~$f_{ij}$ and the difference between any two transition frequencies, which is one of the requirements detailed above.

\subsubsection{Chern insulator on the Lieb lattice}
\label{sec:ChernLieb}
\begin{figure*}[ht!]
\includegraphics[trim = 0mm 0mm 0mm 0mm, clip, width=14cm]{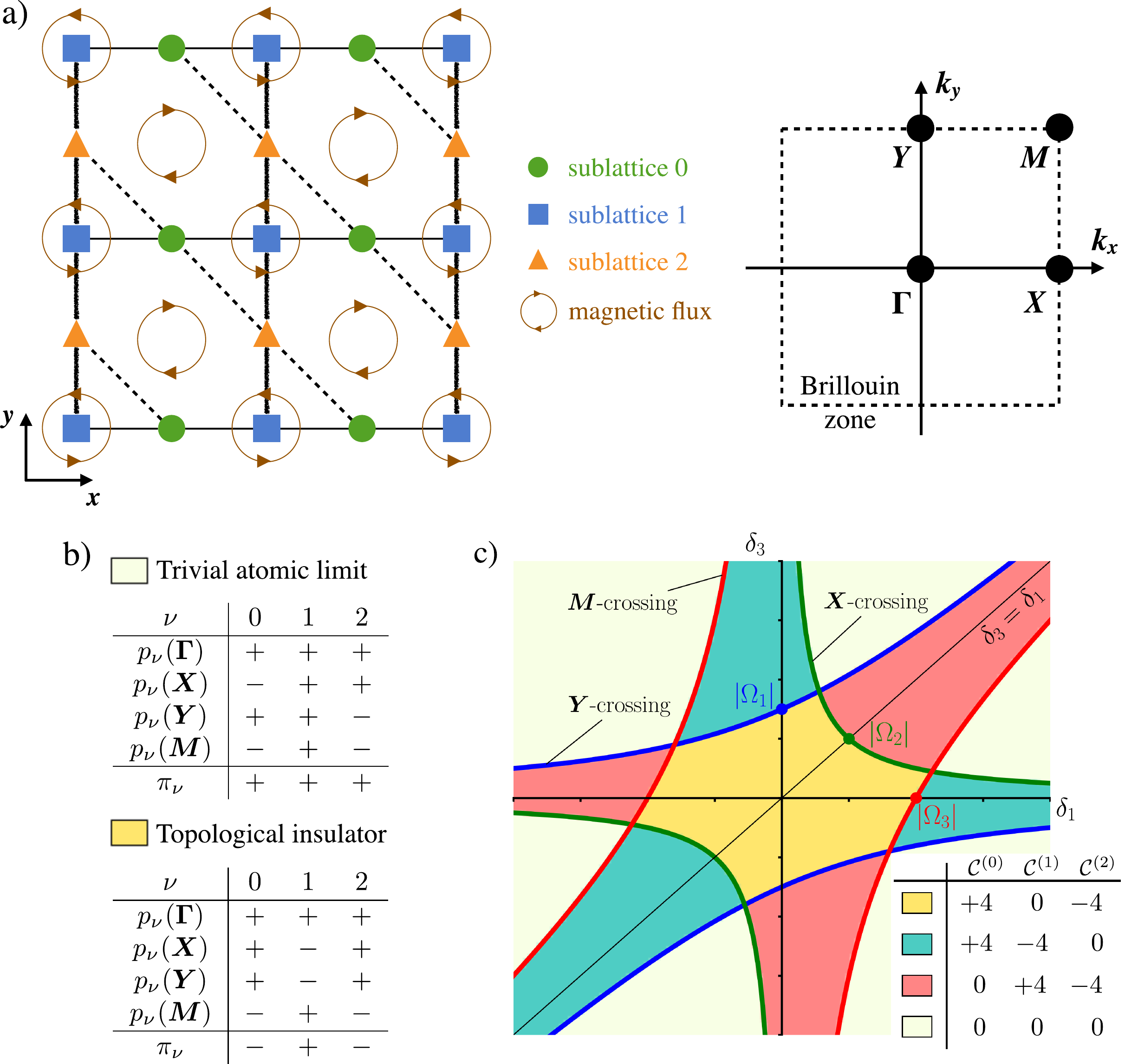}
\caption{%
(a) Schematic representations of a Haldane model on the Lieb lattice and its square Brillouin zone with the four inversion-invariant momenta. The sublattices 0, 1, and 2 respectively have on-site potential energies 0, $\delta_1$, and $\delta_3$. The regular and bold black lines represent the nearest-neighbor tight-binding amplitudes $\Omega_1$ and $\Omega_2$. The dashed black lines depict the next-nearest-neighbor coupling $\Omega_3$.
(b) Two allowed configurations of the band parities $p_\nu$ at the inversion-invariant momenta. A trivial insulator can be adiabatically connected to an atomic limit, which is necessarily characterized here by positive parity products $\pi_\nu=+1$. A band insulator exhibiting negative parity products $\pi_\nu=-1$ is therefore topological. Enumerating all the configurations of the parity eigenvalues allowed by the model parameters leads to the topological phase diagram in panel (c) (see Appendix~\ref{annexe:LiebLatticeAnalysis}).
(c)
Phase diagram of the qutrit model. Each colored region corresponds to a set of band Chern numbers. On the boundaries the gap closes for at least one combination of phases $\phi_1$ and $\phi_2$.
}
\label{fig:LiebModel}
\end{figure*}

In this section, we show that the fluxonium qutrit
emulates in time the physics in momentum space of a Chern insulator on a Lieb lattice, which is illustrated in Fig.~\ref{fig:LiebModel}(a). This proposal thus provides a missing implementation of a new three level topological model, which in particular supersedes a recent proposal based on  molecular enantiomers~\cite{schwennicke2021enantioselective}. 
In this quantum simulation correspondence, the drive detunings $\delta_{1}$ and $\delta_{3}$  correspond to onsite potential energies on the Lieb lattice, while 
the drive amplitudes $\Omega_{1}$ and $\Omega_{2}$ mimic nearest-neighbour tight-binding amplitudes, and $\Omega_{3}$ simulates a 
next-nearest-neighbour coupling along one diagonal direction. 
The relative $\pi$ phase between the $\Omega_{1}, \Omega_{2}$ and $\Omega_{3}$
terms in Eq. (\ref{eq:Hphi1phi2phi3}) originates from 
a periodic pattern of staggered magnetic fluxes, as shown in Fig.~\ref{fig:LiebModel}(a).
These fluxes break the time-reversal symmetry of the tight-binding model, while preserving the translation invariance of the Bravais lattice.
 
We thus end up with a generalization of the celebrated 
Haldane's model \cite{Haldane:1988} but on the Lieb instead of the honeycomb lattice. 
In the sublattice basis $(0,1,2)$ [see Fig.~\ref{fig:LiebModel}(a)], the corresponding Bloch 
Hamiltonian is 
\begin{align}\label{HLieb}
&H(\boldsymbol{k}) = \\
&\hbar\left(
\begin{array}{ccc}
0 & \Omega_{1}\cos\left(\frac{k_{x}}{2}\right) & -i\Omega_{3}\cos\left(\frac{k_{x}-k_{y}}{2}\right) \\
\Omega_{1}\cos\left(\frac{k_{x}}{2}\right) & \delta_{1} & \Omega_{2}\cos\left(\frac{k_{y}}{2}\right) \\
i\Omega_{3}\cos\left(\frac{k_{x}-k_{y}}{2}\right) & \Omega_{2}\cos\left(\frac{k_{y}}{2}\right) & \delta_{3} \\
\end{array}
\right). \notag 
\end{align}
The pseudo-momentum $\boldsymbol{k}$ is dimensionless, corresponding to a lattice constant chosen as a length unit. 
The qutrit Hamiltonian (\ref{eq:Hphi1phi2phi3}) is then recovered through the substitutions $k_{x}\rightarrow 2\phi_1$ and $k_{y}\rightarrow 2\phi_2$.
Note that our model also captures the dynamics of other quantum systems such as spin chains~\cite{vepsalainen2020simulating}.

We now aim at determining the ranges of drive parameters $\Omega_i, \delta_i$ 
that lead to topologically nontrivial band structures for the model of Eq.~(\ref{HLieb}). 
By definition, a topological band structures cannot be smoothly deformed to that of an atomic limit \cite{Cano2018building}. In contrast, a trivial band structure admits some band representations of the crystal space group on a basis of symmetric localized orbitals \cite{zak1980symmetry,zak1981band,zak1982band,michel2001elementary}. An efficient strategy to detect topological band structures then consists of enumerating all possible band representations of a space group and identifying band structures that do not support such representations. This strategy lies at the heart of the recent paradigm of Topological Quantum Chemistry and led to the predictions of exhaustive catalogues of topological materials \cite{bradlyn2017topological,po2017symmetry,vergniory2019complete,tang2019comprehensive,zhang2019catalogue}.
We can use this methodology to efficiently determine the phase diagram of model \eqref{HLieb}.
 
We first determine the band representations of the Lieb lattice in Fig.~\ref{fig:LiebModel}(a) for the atomic limit $\Omega_{1,2,3}=0$. For non-degenerate onsite energies $\delta_{1,3}\neq0$ and $\delta_{1}\neq\delta_{3}$, and in the presence of staggered magnetic fluxes, the lattice only has inversion symmetry and belongs to the wallpaper space group $p2$. 
The three orbitals occupy the maximal Wyckoff positions $q_{0}=(1/2,0)$, $q_{1}=(0,0)$, and $q_{2}=(0,1/2)$ in the primitive unit cell. 
Their elementary band representations are determined from the band parities $p_\nu=\pm1$ --- i.e. eigenvalues of the parity operator --- at the inversion-invariant momenta in the Brillouin zone depicted in Fig.~\ref{fig:LiebModel}(a) \cite{Cano2021Band}. 
It leads to the band representations summarised in the top panel in Fig.~\ref{fig:LiebModel}(b). The parity product
\begin{align}
\pi_{\nu} = p_{\nu}(\boldsymbol{\Gamma})p_{\nu}(\boldsymbol{X})p_{\nu}(\boldsymbol{Y})p_{\nu}(\boldsymbol{M})
\end{align}
of each band $\nu$ is always positive in the atomic limit. Therefore, band structures with negative parity products fall outside these band representations and are topological.

Away from the atomic limit, {\it i.e.} for $\Omega_{1,2,3}\neq0$, we enumerate all the possible parity configurations of the band structure $H(\boldsymbol{k})$ at the inversion-invariant momenta (see Appendix~\ref{annexe:LiebLatticeAnalysis} and Fig.~\ref{NonBandRepresentations}). 
Each configuration leads to a colored region in the $\delta_{1}\delta_{3}$-plane in Fig.~\ref{fig:LiebModel}(c). 
The ivory colored regions correspond to the parity configurations of the atomic limit in Fig.~\ref{fig:LiebModel}(b). 
Thus, they describe trivial band insulators. 
In contrast, we find that the red, green, and yellow regions exhibit negative parity products, thus characterizing topological insulators. 
As an illustration, the bottom panel in Fig.~\ref{fig:LiebModel}(b) specifies the parity configuration of the yellow region, where $\pi_\nu=-1$ for $\nu=0$ and $\nu=2$. 
It shows that the change of parity products between the trivial atomic limit and the topological insulator requires the bands $\nu=1$ and $\nu=0$ ($\nu=2$) to switch parities at point $\boldsymbol{X}$ ($\boldsymbol{Y}$) of the Brillouin zone. 
More generally, a parity switch cannot occur continuously and requires the band gap to close at one 
of the inversion-invariant momentum (see Appendix~\ref{annexe:LiebLatticeAnalysis}). 
 
This provides a very efficient determination of the phase diagram of Fig.~\ref{fig:LiebModel}(c) 
by considering the gap closing at these inversion-invariant momenta as a function of the 
$\Omega_i$ and $\delta_i$. 
Such band crossings mark the borders between topologically distinct regions in the phase diagram represented in Fig.~\ref{fig:LiebModel}(c), where from now on we use the shorthand notation for the Chern number 
$\Chern^{(\nu)} = \Chern^{(\nu)}_{12}$. 
One can further show that this  Chern number is non zero when the parity product is negative~\cite{Hugues:2011,Fang:2012}. 
Since the qutrit Hamiltonian in Eq. (\ref{eq:Hphi1phi2phi3}) is recovered via the substitutions $k_{x,y}\rightarrow 2\phi_{1,2}$, the Chern number for the fluxonium qutrit is four times larger than for the Lieb insulator, hence the values summarized in the table in Fig.~\ref{fig:LiebModel}(c).

\begin{figure*}[!t]
\includegraphics[trim = 0mm 0mm 0mm 0mm,clip,width=15cm]{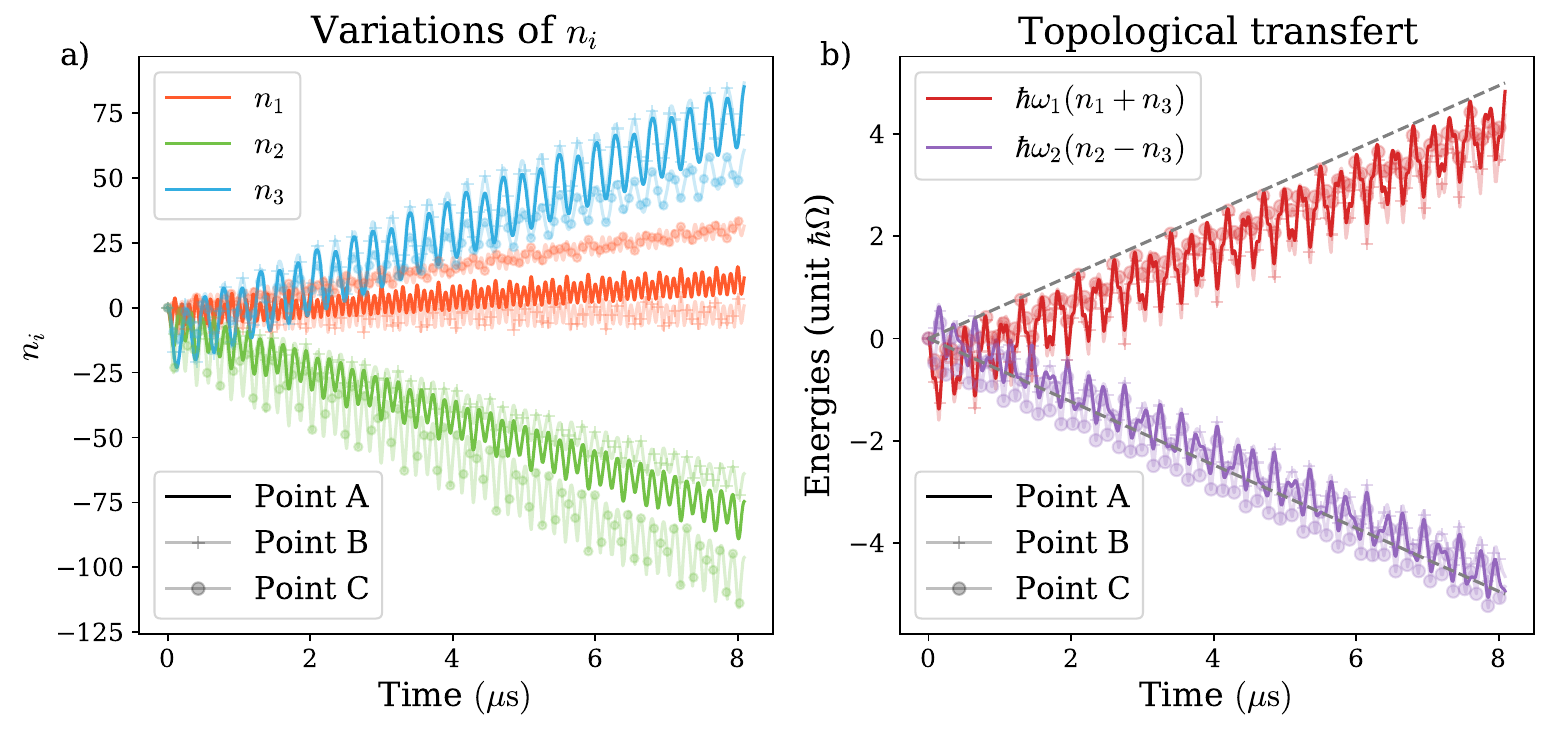}
\caption{
Pumping by a qutrit initialized in its ground state. 
(a) Filling of the different modes $n_i$ as a function of time, for three different set of pump parameters corresponding to points A, B and C in Fig.~\ref{fig:TopTransition}(a). 
The driving amplitudes are chosen all equals $\Omega_{1,2,3}=\Omega$. 
The filling rate are found to depend on the parameters of the pump. 
(b) The two different combinations of fillings display the topological power~$\hbar\omega_1\omega_2 \Chern^{(0)}/2\pi$ (grey dashed lines), invariant on the region of stability of the phase diagram,
where $\Chern^{(0)}=4$ is the Chern number of the ground state for all parameter sets A,B,C. 
}
\label{fig:pumpingSeveralPoints}
\end{figure*}

\subsubsection{Topological power transfer between three modes}\label{sec:dynamicsModes}

We now discuss how the topological nature of the qutrit pump manifests itself in filling rate of the
three modes. 
The dynamical system consist of three classical modes described by the classical phases~$\phi_i$ conjugated to~$n_i$
coupled to the qutrit through the Hamiltonian~\eqref{eq:Hphi1phi2phi3} in the rotating frame. 
The equations of dynamics have the same form in the rotating frame
$\dot{n}_i = -\frac{1}{\hbar}\bra{\Psi} \partial_{\phi_i} \tilde H  \ket{\Psi}, 
i=1,2,3$
where the dynamics of the qutrit state~$\ket{\Psi(t)}$ in the rotating frame is governed by the Hamiltonian~\eqref{eq:Hphi1phi2phi3},  
see Appendix~\ref{annexe:dynamicsRotatingFrame} for details.
 
As said above, the frequencies of amplitude modulation satisfies~$\omega_3=\omega_1-\omega_2$ such that
$\phi_{III}=\phi_1-\phi_2-\phi_3$ is a constant of motion, so
we can keep $\phi_1 - \phi_2 - \phi_3 = 0$ at all time.
We consider the following canonical transformation
$n_I = n_1 + n_3 , n_{II}  = n_2 - n_3, 
\phi_{I}   = \phi_1, \phi_{II}  = \phi_2$. 
The dynamics of~$n_I$ and~$n_{II}$ is given by
\begin{subequations}\label{eq:nI_nII_dynamics}
\begin{align}
    \dot{n}_1 + \dot{n}_3 =
    -\frac{1}{\hbar}\frac{\partial E_\nu}{\partial \phi_1} + \Berry^{(\nu)}_{\phi_1 \phi_2}\omega_2 \\
    \dot n_2 - \dot n_3 = 
    -\frac{1}{\hbar}\frac{\partial E_\nu}{\partial \phi_2} - \Berry^{(\nu)}_{\phi_1 \phi_2}\omega_1
\end{align}
\end{subequations}
with~$E_\nu$ the energy and~$\Berry^{(\nu)}$ the Berry curvature of the band~$\nu$ of the Hamiltonian~$H(\phi_1,\phi_2)\equiv\tilde H(\phi_1, \phi_2, \phi_1-\phi_2)$ 
in which the qutrit is initially prepared (see Appendix~\ref{annexe:dynamicsRotatingFrame}).
Then, the topological power transfer between the modes~$1,2,3$ is 
\begin{equation}
    \hbar\omega_1 \langle  \dot n_1 + \dot n_3\rangle_t = \hbar\frac{\Chern^{(\nu)}}{2\pi}\omega_1\omega_2
    = -\hbar\omega_2 
    \langle  \dot n_2 - \dot n_3\rangle_t,
    \label{Eq:tocheckexperimentally0}
\end{equation}
with~$\Chern^{(\nu)}$ the Chern number of the band~$\nu$ of the Hamiltonian.
 
From a more experimental point of view, the power transfer to demonstrate is
\begin{equation}
    \frac{\Delta S_1}{h f_1} + \frac{\Delta S_3}{h f_3} = \frac{\Chern^{(\nu)}}{2\pi}\omega_2\textrm{ and }   \frac{\Delta S_2}{h f_2} - \frac{\Delta S_3}{h f_3} = -\frac{\Chern^{(\nu)}}{2\pi}\omega_1.\label{Eq:tocheckexperimentally}
\end{equation}
In order to get a sense of how feasible this measurement is, let us set some possible figures for the experiment that fulfill the criterion discussed earlier. The fluxonium frequencies could be set to $f_{01}=4~\mathrm{GHz}$, $f_{12}=6~\mathrm{GHz}$ and $f_{02}=10~\mathrm{GHz}$. It is then possible to drive the transitions with $\Omega_{1,2,3}/2\pi=100~\mathrm{MHz}$ and a similar range of variation for the detunings. The modulation frequencies could then be $\omega_{1}/2\pi=5~\mathrm{MHz}$, $\omega_{2}/2\pi\simeq3~\mathrm{MHz}$. 
From the simulations below, we see that the topological power transfer can be resolved in about 30 periods $2\pi/\omega_{i}$, which is a few $\mu\mathrm{s}$. This is well below the typical decoherence times of fluxonium qubits, which will thus not limit the dynamics of the system during the measurement. Verifying Eq.~(\ref{Eq:tocheckexperimentally}) thus requires to measure instantaneous powers in the range of~$h f_{i}\omega_{i}$. This corresponds to a power of several dozens of aW, which is a level of precision that is now routinely reached experimentally~\cite{cottet2017observing,ronzani2018tunable,kokkoniemi2020bolometer}.

\begin{figure*}[t]
\includegraphics[trim = 0mm 0mm 0mm 0mm,clip,width=17cm]{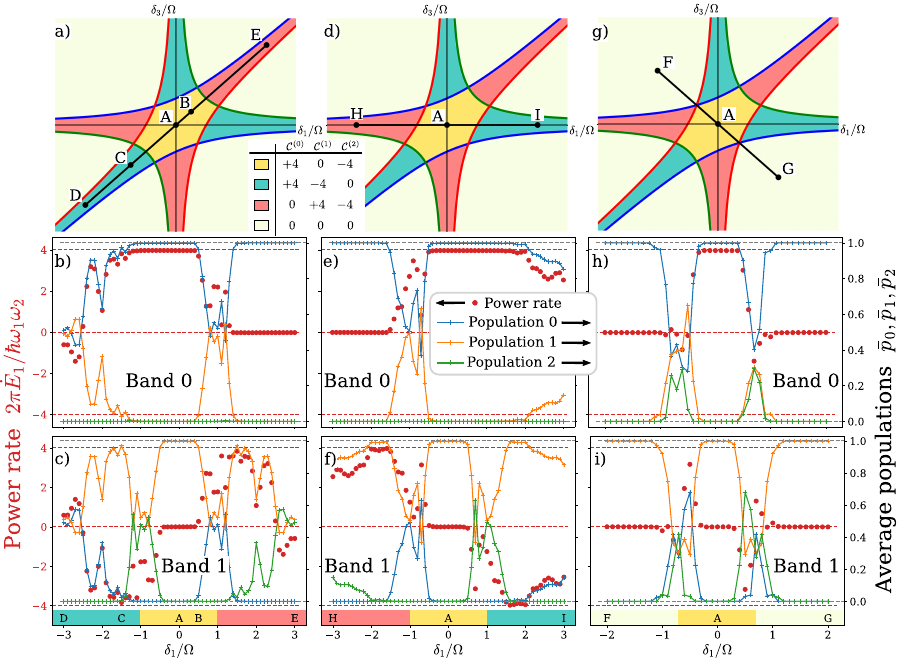}
\caption{
Topological transitions for bands 0 and 1 in the case of same drive amplitudes~$\Omega_{1,2,3}=\Omega$.
For each working point of the phase diagram, we fit the time evolution of the energy~$E_1=\hbar\omega_1(n_1+n_3)$ by a line on a time~$\Delta t=8\mu\mathrm{s}$ to construct the reduced power rate~$2\pi\dot E_1/\hbar\omega_1\omega_2$ (red dots) 
when the qutrit is initialized in band~0, panels~[(b),(e),(h)], or in band~1, panels~[(c),(f),(i)]. 
The average populations of the qutrit~$\bar p_\mu=\frac{1}{\Delta t}\int_0^{\Delta t}|\braket{\psi_\mu(t)}{\Psi(t)}|^2\dd t$ are displayed to evaluate adiabaticity.
[(a)-(c)] Transition line~DE, with a transition between bands~1 and~2 at~$\delta_1/\Omega=-1$, and between bands~0 and~1 at~$\delta_1/\Omega=1$. 
[(d)-(f)] Transition line~HI, with a transition between bands~0 and~1 at~$\delta_1/\Omega=-1$, and between bands~1 and~2 at~$\delta_1/\Omega=1$.
[(g)-(i)] Transition line~FG, with a transition between the three bands at~$\delta_1/\Omega=\pm\sqrt{2}/2$.
}
\label{fig:TopTransition}
\end{figure*}

\subsection{Numerical analysis of topological pumping}

\subsubsection{Topological power transfer}
In order to perform numerical simulations of the proposed experiment, 
we solve the time-dependent Schr{\"o}dinger equation of the qutrit
under the rotating wave approximation~\eqref{eq:Hphi1phi2phi3}. 
We first determine the optimal parameters for the pump: the stability of the 
adiabaticity evolution requires the largest gap. 
This is reached at the resonant drive
~$\delta_1=\delta_3=0$, point~A  in the phase diagram in Fig. \ref{fig:TopTransition}(a),
and with equal driving amplitudes $\Omega_{1,2,3}=\Omega$, resulting in a gap~$0.87~\hbar\Omega$. 
In these conditions, from the analysis of Fig.~\ref{fig:LiebModel}(c) 
both bands~0 and~2 have non-zero Chern number~$\Chern=\pm 4$, whereas the band~1 is topologically trivial with~$\Chern=0$.
Therefore the two lowest energy states do not constitute an effective topological qubit, leading to different dynamics than for a conventional $2$~level pump as we will see below.  
To ensure an ergodic exploration of the classical configuration space of the pump, we choose the ratio between phases frequencies~$\omega_1 / \omega_2 =(1+\sqrt{5})/2$.

Keeping parameters of the pump to point~A in the phase diagram in Fig.~\ref{fig:TopTransition}(a), we initialize the qutrit at $t=0$ in its 
ground state. The evolution of the filling $n_i$ of each individual modes is represented in 
 Fig.~\ref{fig:pumpingSeveralPoints}(a), and varies linearly in time. However the filling rate $\dot{n}_i$ are not 
set by the topological nature of the pump, and depend on the precise values of the pump parameters: changing slightly 
these from point A to point B in the same topological region of Fig.~\ref{fig:TopTransition}(a) leads to different filling rates, 
as seen in Fig.~\ref{fig:pumpingSeveralPoints}(a). 
On the other hand, the topological power transfer defined in Eq.~(\ref{Eq:tocheckexperimentally0}) is insensitive 
of the precise values of the pump parameters, as shown in Fig.~\ref{fig:pumpingSeveralPoints}(b).
 
Besides the linear topological evolution in time, this power transfer displays temporal fluctuations which 
have two different origins as deduced from Eq.~\eqref{eq:nI_nII_dynamics}: a 
dominant spectral term, corresponding to variation of the energy $E_0$ of the qutrit, 
and a geometrical contribution originating from fluctuations of the Berry curvature around its topologically quantized average value (see Appendix~\ref{annexe:TempFluctuations}).
Thus the order of magnitude of the correlation timescale of these fluctuations correspond to the  period of the drive. 
A reasonable requirement to detect the topological power transfer is to average it over~$30$ such independent fluctuations, leading to a measurement time of 8~$\mu$s, as announced in the previous section.

\subsubsection{Numerical detection of topological transitions}
Having established that the average topological power transfer gives access to the Chern number of the band in which the qutrit was initialized, 
we now address the detection of the topological phase transitions of Fig.~\ref{fig:LiebModel}(c) when the  
detuning parameters $\delta_1,\delta_2$ are varied. 
The richness of the adiabatic dynamics of the qutrit pumps require different experimental protocols adjusted to each phase transition. 
For example, sets A and C of parameters in  Fig.~\ref{fig:TopTransition}(a) lead to exactly the same topological power rate for a qutrit initialized in the ground state, as shown in  Fig.~\ref{fig:pumpingSeveralPoints}(b), 
but corresponds to a different topological qutrit phase.
Indeed, the corresponding qutrit phases differ by the topological nature of the 
excited bands 1 and 2, while the nature of the ground state is unchanged. 
Thus detecting this particular transition requires an initialization of the qutrit
in the first excited state 1.
 
To detect all transitions, we monitor the evolution of the pumps with both initialization in the 0 and 1 states. 
Figure \ref{fig:TopTransition} displays the resulting 
power rate, determined by a linear fit 
using Eq.~\eqref{Eq:tocheckexperimentally0}, 
as well as the  average populations~
$\bar p_\mu=\frac{1}{\Delta t}\int_0^{\Delta t} |\braket{\psi_\mu(t)}{\Psi(t)}|^2\dd t$ 
of the qutrit state~$\ket{\Psi(t)}$ on the three instantaneous eigenstates~$\ket{\psi_\mu(t)}$. 
Figures \ref{fig:TopTransition}(b,e,h) correspond to a pump with qutrit initialized in band~0, while Fig.~\ref{fig:TopTransition}(c,f,i) correspond to an initial preparation in band~1.

Along the lines DE and HI in Fig.~\ref{fig:TopTransition}(a,d), the 
topological transitions occur between two bands only, whereas along the line FG in Fig.~\ref{fig:TopTransition}(g) the gaps between all three bands close at the transitions. 
Along any line, the transition is detected by the evolution of the populations. 
Moreover, a quantized power transfer in a given state appears as a direct test of the adiabatic nature of the evolution, related to the distance to the transitions. 
In that respect, optimal choice of parameters for the pump correspond to  point~A in the yellow topological phase
of the phase diagram, in which the bands~0 and~2 are non-trivial and spectrally separated by a trivial band~1  and thus generically separated from a trivial phase by two transitions. 
Far from any transition near point~A, 
the instantaneous energy separation with different eigenstates  is large, resulting in an adiabatic evolution: 
the average population of the qutrit in the initial band remains close to~$1$ and the power rate is quantized and set by the Chern number of the band.

In the other topological states of the qutrit, the effects of non-adiabaticity are manifest, resulting from shorter distances to phase transitions and thus small gaps. 
For example along line~DE for a qutrit prepared in band~1,  while 
the Chern number~$\Chern^{(1)}$ takes values~$-4$ and~+4 for respectively~$\delta_1/\Omega<-1$ and~$\delta_1/\Omega>1$, 
the qutrit does not evolve adiabatically and the dynamics of the classical variables~\eqref{eq:nI_nII_dynamics} must be corrected, leading to an unquantized power transfer 
shown in Fig.~\ref{fig:TopTransition}(c). 
Similarly in Fig.~\ref{fig:TopTransition}(b) the Chern number
$+4$ of band~0 for $\delta_1/\Omega<1$ manifests itself as a pleateau of power rate for a reduced set of parameters  for~$-1.5<\delta_1/\Omega<0.5$ (between points~C and~B).

\begin{figure*}[t]
\includegraphics[trim = 0mm 0mm 0mm 0mm,clip,width=16cm]{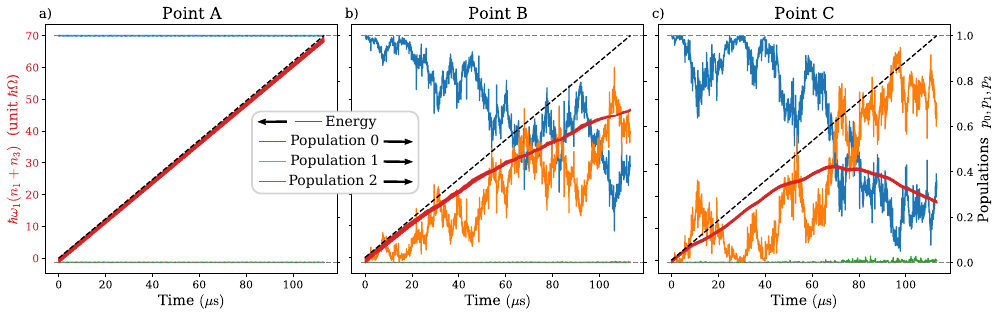}
\caption{Non-adiabatic effect at longer time. 
The time evolution for the qutrit prepared in band~0 of the topological energy combination~$\hbar\omega_1(n_1+n_3)$ and qutrit populations~$p_\mu(t)$ in the instantaneous eigenstates are displayed for three working point~A,~B and~C in the phase diagram of Fig.~\ref{fig:TopTransition}(a).
(a) Point~A, case of resonance~$\delta_1=\delta_3=0$ where the evolution of the qutrit remains adiabatic during~200 periods so the pumping rate is stable.
(b) Point~B, limit case beyond which pumping is no longer quantized on Fig.~\ref{fig:TopTransition}(b), where the evolution of the qutrit is no longer adiabatic after approximately $8~\mu\mathrm{s}$ so the pumping rate is no longer quantized.
(c) Point~C, other limit case beyond which pumping is no longer quantized on Fig.~\ref{fig:TopTransition}(b), where in this phase the Chern number of band~1 is opposite to band~0 so we see pumping in the other direction when band~1 is mainly populated.
Downsampling of data has been applied for clarity  of presentation.
}
\label{fig:pumpingLongTime}
\end{figure*}
\subsubsection{Non-adiabatic pumping}

The importance of the non-adiabatic effects can be anticipated from the estimation of the time of validity of the adiabatic approximation~$\tau_\mathrm{adiab}$ introduced in Eq.~\eqref{eq:adiabTime}.
In this expression, the maximum of the adiabatic parameter~$\epsilon_{\mu\nu}$ is taken on all the values of phases~$\phi_1$ and~$\phi_2$, and the mean free time~$\tau_\mathrm{mft}$ is the order of magnitude of the phases periodicity, we take~$\tau_\mathrm{mft}=2\pi/\omega_2\sim 0.3~\mu\mathrm{s}$.
For the point A,~B and~C of the phase diagram, the Landau-Zener collision time~$\tau_{01}^\mathrm{col}$ associated to the transition between states~0 and~1 is one order of magnitude below this mean free time, which is consistent with the derivation of the adiabatic time~$\tau_\mathrm{adiab}$ discussed in Sec.~\ref{sec:conditionAdiab}.
 
For the point~A of maximal stability with a preparation in band~0, we get~$\tau_\mathrm{adiab}^\mathrm{A}\sim 150~\mathrm{ms}$ so the non-adiabatic effects will not be limiting for experiments with these parameters values.
This is illustrated in Fig.~\ref{fig:pumpingLongTime}(a), where the evolution for~100~$\mu\mathrm{s}$ of the energy~$\hbar\omega_1 (n_1+n_3)$ and the populations~$p_\mu(t)=|\braket{\psi_\mu(t)}{\Psi(t)}|^2$
of the qutrit state~$\ket{\Psi(t)}$ on the three instantaneous eigenstates~$\ket{\psi_\mu(t)}$ are displayed.
The qutrit stays in the ground state with~$p_0(t)>0.997$, and the energy is transferred at the topologically quantized rate.
 
Figure~\ref{fig:pumpingLongTime}(b) corresponds to point~B
closer to the topological transition line towards the phase where~$\Chern^{(0)}=0$.
The estimated time of adiabaticity for band~0 is~$\tau_\mathrm{adiab}^\mathrm{B}\sim 6~\mu\mathrm{s}$. After this typical time, the population on state~1 exceeds~0.1, in agreement with the definition of~$\tau_\mathrm{adiab}$, and the pumping rate deviate from its topologically quantized value.
Figure~\ref{fig:pumpingLongTime}(c) corresponds to point~C, where we crossed 
a different topological transition line, 
where the ground state remains topological but the first excited state  switches from trivial to non-trivial with Chern number~$\Chern^{(1)}=-4$.
We compute here~$\tau_\mathrm{adiab}^\mathrm{C}\sim 8~\mu\mathrm{s}$
in agreement with the observed deviations from the adiabatic evolution.
At longer times, about~70~$\mu$s, the first excited state is mostly populated and 
the energy pumping is reversed, manifesting the associated change of Chern number.

\section{Discussion}

We have proposed an experiment that is able to observe a topologically protected power exchange. The topological properties appear in the measured incoming and outgoing energy flows that drive a quantum system. Our general description of topological quantum pumps includes the classical degrees of freedom that carry these flows. Considering a qutrit instead of a qubit is key for two aspects. First, it solves the tremendous challenge to probe the power flows that drive a two level topological pump. Second, owing to its richer
dynamics, which simulates the topological band properties of a $3$-band Chern model, it gives access to various protocols of pumping which can be freely chosen by setting the initial state of the qutrit. 
 
Besides the fascinating perspective to actually measure this topological power transfer, 
such a system also opens the paths to the study of the interplay between decoherence and the topological adiabatic evolution of the quantum system. Our framework raises two natural questions.
Topological pumping requires an initial correlation between the qutrit and the classical modes. Then how long does the pumping survive in presence of qutrit decoherence? 
Besides, could we revert the perspective and use the developed framework and measurable pumping rate as a tool to characterize the correlations between the quantum system and the classical modes. 
 
Finally, let us stress that
while we have proposed an experiment demonstrating the topologically protected transfer of microwave power using a superconducting circuit, our general framework can be applied to any quantum system and its driving environment such as cold atoms, mechanical oscillators or polaritons. 
Symmetries are essential to classify topological matter and in particular topological pumps~\cite{meidan2011topological}. The enforcement of these symmetries to protected topological pumping
 in these different systems is a stimulating perspective.

\begin{acknowledgments} 
This research was supported by IDEX Lyon project ToRe (Contract No. ANR-16-IDEX-0005). C.D. also acknowledges the supports of Idex Bordeaux (Maesim Risky project of the LAPHIA Programme) and Quantum Matter Bordeaux. We are grateful to Anton Akhmerov, Antoine Essig, Leonid Glazman, S\'ebastien Jezouin and Christophe Mora for insightful discussions.
\end{acknowledgments}

\bibliography{main}

\newpage 
\appendix

\section{First-order adiabatic evolution}\label{annexe:AdiabEvol}
For the sake of completeness we detail here the derivation of the first-order adiabatic evolution usually derived in different manner in the literature, see for example~\cite{messiah1962quantum, thouless1983quantization, aharonov1987phase,Rigolin:2008,gritsev2012dynamical}.
The quantum Hamiltonian~$H(t)$ depends continuously on time through the set of variables~$\{q_{\alpha}(t)\}$. The time-dependent Schr\"odinger equation is
\begin{align}\label{TDSE}
    i\hbar \frac{\dd}{\dd t}|\Psi(t)\rangle = H(t)|\Psi(t)\rangle ~.
\end{align}
We look for a quantum state of the form
\begin{align}\label{TDState}
    |\Psi (t)\rangle=\sum_\nu a_\nu(t) |\psi_\nu(t)\rangle ~,
\end{align}
where $|\psi_\nu(t)\rangle$ are the instantaneous eigenstates of $H(t)$ and satisfy
\begin{align}
    H(t) |\psi_\nu(t)\rangle = E_\nu(t) |\psi_\nu(t)\rangle ~.
\end{align}
This leads to the equation of motion
\begin{multline}\label{EquationOfMotion}
    \dot a_\nu(t) = i \left[ i \langle \psi_\nu (t) | \frac{\dd}{\dd t} | \psi_\nu(t)\rangle - \frac{1}{\hbar}E_\nu(t) \right] a_\nu(t) 
    \\- \sum_{\mu\neq\nu} \langle \psi_\nu(t) | \frac{\dd}{\dd t} | \psi_\mu(t)\rangle a_\mu(t) ~.
\end{multline}
We can get rid of the first term in the right-hand side by introducing the phases
\begin{align}
    \Delta_\nu(t) =-\frac{1}{\hbar}\int_0^t \dd t'~ E_\nu(t')
\end{align}
and
\begin{align}
    \Gamma_\nu(t) =\int_0^t \dd t'~ \bra{\psi_\nu(t')} i \frac{\dd}{\dd t'} \ket{\psi_\nu(t')},
\end{align}
and making the substitution $a_\nu(t)=e^{i \Delta_\nu(t)+i\Gamma_\nu(t)}\tilde{a}_\nu(t)$.
The time integration of Eq.~(\ref{EquationOfMotion}) then results in
\begin{multline}\label{TimeIntegral}
    \tilde{a}_\nu(\tau) - \tilde{a}_\nu(0)
    = \\- \sum_{\mu\neq\nu} \int_0^\tau d\tau' A_{\mu\nu}(\tau')
    e^{i\Gamma_{\mu\nu}(\tau')}
    e^{\frac{1}{\eta}\int_0^{\tau'} d\tau'' \Delta_{\mu\nu}(\tau'')}
\end{multline}
where $A_{\mu\nu}(\tau)=\langle \psi_\nu(\tau) | \partial_\tau \psi_\mu(\tau)\rangle \tilde{a}_\mu(\tau)$, $\Gamma_{\mu\nu}(\tau)=\Gamma_{\mu}(\tau)-\Gamma_{\nu}(\tau)$, $\Delta_{\mu\nu}(\tau)=-i(E_\mu(\tau)-E_\nu(\tau))/(\hbar\Delta)$, and $\Delta = \min_{\mu,t}|E_\mu(t)-E_\nu(t)|/\hbar$ is assumed to be non zero. Here, we have also introduced a dimensionless ``time" $\tau=\eta \Delta t$, where $\eta$ is a small parameter. 
We can evaluate the time integral in Eq.~(\ref{TimeIntegral})
by integrating by part on the exponential term
\begin{equation}
    e^{\frac{1}{\eta}\int_0^{\tau'} d\tau'' \Delta_{\mu\nu}(\tau'')}
    =
    \frac{\eta}{\Delta_{\mu\nu}(\tau')}\frac{\dd}{\dd\tau'}\left[e^{\frac{1}{\eta}\int_0^{\tau'} d\tau'' \Delta_{\mu\nu}(\tau'')}\right].
\end{equation}
The first-order correction in the adiabatic limit results in
\begin{multline}
    \tilde{a}_\nu(\tau) - \tilde{a}_\nu(0) = \\
    - i\hbar\eta\Delta \sum_{\mu\neq\nu}
    \frac{\langle \psi_\nu(\tau) | \partial_\tau \psi_\mu(\tau)\rangle}{E_\mu(\tau)-E_\nu(\tau)} \tilde{a}
    _\mu(\tau)e^{i\Gamma_{\mu\nu}(\tau)}e^{\frac{1}{\eta}\int_0^{\tau} d\tau' \Delta_{\mu\nu}(\tau')}  \\
    + i\hbar\eta\Delta \sum_{\mu\neq\nu}
    \frac{\langle \psi_\nu(0) | \partial_\tau \psi_\mu(0)\rangle}{E_\mu(0)-E_\nu(0)} \tilde{a}_\mu(0)
    +\mathcal{O}(\eta^2).
    \label{eq:firstOrderEqAtilde}
\end{multline}
If the quantum system is initially prepared in the instantaneous eigenstate $|\psi_\nu\rangle$ --- i.e. $a_\nu(0)=1$ and $a_{\mu\neq\nu}(0)=0$ --- we find the time dependent state
\begin{widetext}
\begin{multline}
    \ket{\Psi (t)} = 
    e^{i\Gamma_\nu(t)+i\Delta_\nu(t)} 
    \left(
        \ket{\psi_\nu(t)}  - i \hbar \sum_{\mu\neq \nu} 
        \frac{\bra{\psi_\mu(t)}  \frac{\dd}{\dd t} \ket{\psi_\nu(t)}}
         {E_{\nu}(t)-E_{\mu}(t)} \ket{\psi_\mu(t)}
    \right)
    \\
     +i \hbar \sum_{\mu\neq\nu}  e^{i\Gamma_\mu(t)+i\Delta_\mu(t)} 
      \left[ \frac{\bra{\psi_\mu(t)}  \frac{\dd}{\dd t} \ket{\psi_\nu(t)}}
                  {E_{\nu}(t)-E_{\mu}(t)} 
      \right]_{t=0} \ket{\psi_\mu(t)}
     +\mathcal{O}(\eta^2)
    \label{eq:PsiAdiab}
\end{multline}
\end{widetext}
where the components on~$\ket{\psi_{\mu\neq\nu}}$ are first order terms in~$\eta$.
Using the identity $\bra{\psi_\mu(t)} \frac{\dd}{\dd t} \ket{\psi_\nu(t)}=\bra{\psi_\mu}\frac{\dd H}{\dd t}\ket{\psi_\nu}/(E_\nu-E_\mu)$, the correction at first order in~$\eta$ to the equation of motion of the variable~$p_\alpha$ is
\begin{multline}\label{eq:avCurrent}
    -\bra{\Psi}\frac{\partial H}{\partial q_\alpha}\ket{\Psi} = -\bra{\psi_\nu}\frac{\partial H}{\partial q_\alpha}\ket{\psi_\nu} \\
    +i\hbar\sum_{\mu\neq\nu}
    \frac{
    \bra{\psi_\nu}\frac{\partial H}{\partial q_\alpha}\ket{\psi_\mu}
    \bra{\psi_\mu}\frac{\dd H}{\dd t}\ket{\psi_\nu}}{\left(E_\nu-E_\mu\right)^2}\\
    -i\hbar\sum_{\mu\neq\nu}
    \frac{
    \bra{\psi_\nu}\frac{\dd H}{\dd t}\ket{\psi_\mu}
    \bra{\psi_\mu}\frac{\partial H}{\partial q_\alpha}\ket{\psi_\nu}}{\left(E_\nu-E_\mu\right)^2}
\end{multline}
where the last term of~\eqref{eq:PsiAdiab} induces only first order terms rapidly oscillating at the Bohr frequencies~$(E_\nu-E_\mu)/\hbar$ which has no effect on the dynamics at the time-scale of the slow variables.
We note that this terms are due to the choice of initial condition and could be eliminated if we choose~$a_\nu(0)=1$ 
and~$a_{\mu\neq\nu}(0)=-i\hbar\left[\bra{\psi_\mu(t)}\frac{\dd}{\dd t}\ket{\psi_\nu(t)}/(E_{\nu}(t)-E_{\mu}(t))\right]_{t=0}$.
For the implementation with a superconducting qutrit, a qutrit is fully controllable and any quantum state can be prepared~\cite{bianchetti2010control,peterer2015coherence,ficheux:tel-02098804}.
Using this initial condition, we recover the adiabatic parameter~$\epsilon_{\mu\nu}$ introduce in Sec.~\ref{sec:conditionAdiab} in the population on an excited state~$|\braket{\psi_\mu(t)}{\Psi(t)}|^2=\epsilon_{\mu\nu}(t)^2$.

Returning to Eq.~\eqref{eq:avCurrent},
the time dependence of the Hamiltonian~$H(t)=H(\{q_\beta(t)\})$ is due to the coupling to the classical variables~$q_\beta$, whose classical dynamics is not modified by the coupling to the quantum system, $\dot q_\beta=\dot q_\beta^{(0)}=\frac{\partial \mathcal{H}_\beta}{\partial p_\beta}$.
Thus, using the relation~$\bra{\psi_\nu}\frac{\partial H}{\partial q_\alpha}\ket{\psi_\nu}=\frac{\partial E_\nu}{\partial q_\alpha}$ and the expression of the components of the Berry curvature two form given in~\eqref{eq:curvature}, we obtain for the correction of~$\dot p_\alpha$
\begin{multline}
    -\bra{\Psi}\frac{\partial H}{\partial q_\alpha}\ket{\Psi}= -\frac{\partial E_\nu}{\partial q_\alpha}
    +\hbar \sum_{\beta\neq\alpha} \dot q_\beta^{(0)}F_{q_\alpha q_\beta}.
\end{multline}

\section{Details on the derivation of the Hamiltonian}
\label{annexe:RotFrame}

We detail here the derivation of the Hamiltonian of the qutrit in the rotating frame explained in Sec.~\ref{sec:HamiltonianDerivation}.
The circuit is driven by three drives whose amplitudes
are modulated in time according to the drive Hamiltonian
\begin{equation}\label{eq:QutritLabo}
   H_\mathrm{drive} = \hbar\sum_{i=1}^3 g_i \cos(\phi_i)\cos(\theta_i) \hat N.
\end{equation}
with $\theta_i(t)=2\pi f_i t + \theta_i^0$ the phase of the the electromagnetic field of frequency $f_i$, $\phi_i(t)=\omega_i t$ the phase of the time modulation of the amplitude, and $g_i$ the coupling rates.
In the basis~$(\ket{0}, \ket{1}, \ket{2})$ of the three eigenstates of~$H_\flux$~\eqref{eq:fluxoniumHam} of lowest energy, the diagonal elements of $\hat N$ are null so the Hamiltonian in the laboratory frame~$H_\labo=H_\flux+H_\drive$ has the form:
\begin{multline}\label{eq:Hlabo}
    H_\labo
    = \\
    \hbar\left( 
    \begin{array}{ccc}
	0 & \displaystyle\sum_{i=1}^3 \Omega_{i,01} \cos(\phi_i) \cos(\theta_i) & \displaystyle\sum_{i=1}^3 \Omega_{i,02}\cos(\phi_i) \cos(\theta_i) \\
	c.c. & 2\pi f_{01} & \displaystyle\sum_{i=1}^3 \Omega_{i,12}\cos(\phi_i) \cos(\theta_i) \\
	c.c. & c.c. & 2\pi f_{02}
	\end{array}
    \right)
\end{multline}
where $f_{01}$ and $f_{02}$ are the transition frequencies of~$H_\flux$, and
$\Omega_{i,ab} = g_i \bra{a}\hat N \ket{b}$.
We change of reference frame with the unitary transformation
$U(t) = \textrm{diag} (1, \exp(-i2\pi f_1 t), \exp(-i 2\pi f_3 t) )$.
The frequencies of the three pumps satisfy the constraint
$f_3 = f_1 + f_2$, so the Hamiltonian in the rotating frame is
\begin{widetext}
\begin{align}
    H_\rot
    &= U^\dagger H_\labo U - i\hbar U^\dagger\frac{\dd U}{\dd t}
    = \hbar\left( 
    \begin{array}{ccc}
	0 & e^{-i2\pi f_1 t} \displaystyle\sum_{i=1}^3 \Omega_{i,01}\cos(\phi_i) \cos(\theta_i) & e^{-i2\pi f_3 t}\displaystyle\sum_{i=1}^3 \Omega_{i,02}\cos(\phi_i) \cos(\theta_i) \\
	c.c. & \delta_{1} & e^{-i2\pi f_2 t}\displaystyle\sum_{i=1}^3 \Omega_{i,12}\cos(\phi_i) \cos(\theta_i) \\
	c.c. & c.c. & \delta_{3}
	\end{array}
    \right)
\end{align}
\end{widetext}
with the detuning $\delta_{i}$ given by
\begin{align}
    \delta_{1} &= 2\pi f_{01} - 2\pi f_1 \\
    \delta_{2} &= 2\pi f_{12} - 2\pi f_2 \\
    \delta_{3} &= 2\pi f_{02} - 2\pi f_3 = \delta_1 + \delta_2.
\end{align}
The phases of the drives are given by $\theta_i(t) = 2\pi f_i t + \theta_i^0$.
In the rotating wave approximation, we ignore the terms of the Hamiltonian in the rotating frame which oscillate at the frequency of the drives, so we approximate the Hamiltonian by
\begin{multline}
    \label{eq:QutritRotFrame}
    H_\rot(t) \simeq H(\phi_1,\phi_2,\phi_3) = \\
    \hbar\begin{pmatrix}
    0 & \frac{1}{2}\Omega_{1,01}\cos(\phi_1)e^{i\theta_1^0} & \frac{1}{2}\Omega_{3,02}\cos(\phi_3)e^{i\theta_3^0} \\
	c.c. & \delta_{01} & \frac{1}{2}\Omega_{2,12}\cos(\phi_2)e^{i\theta_2^0} \\
	c.c. & c.c. & \delta_{02}
	\end{pmatrix}
\end{multline}
where the other terms oscillates at the frequency $f_i\pm f_j$. 
We recover the Hamiltonian~\eqref{eq:Hphi1phi2phi3} by noting the drive amplitudes~$\Omega_1 = \frac{1}{2}|\Omega_{1,01}|=\frac{1}{2}g_1 |\bra{0}\hat N\ket{1}|$, $\Omega_2 = \frac{1}{2}|\Omega_{2,12}|=\frac{1}{2}g_2 |\bra{1}\hat N\ket{2}|$ and~$\Omega_3 = \frac{1}{2}|\Omega_{3,02}|=\frac{1}{2}g_3 |\bra{0}\hat N\ket{2}|$, and by choosing the initial pump phases~$\theta_i^0$ to set the complex phase of the couplings at the desired value.
The rotating wave approximation is valid if the drive amplitudes and detunings are much lower than the frequencies~$f_i\pm f_j$ of the oscillating terms, which means that they must be much lower than the difference between any two transition frequencies of the fluxonium as said in the Sec.~\ref{sec:exp}.

\section{Classical variables coupled to the qutrit \label{sec:powerToMeasure}}

The fluxonium is coupled to three drives whose amplitudes are modulated in time.
The coupling Hamiltonian in the laboratory frame is
\begin{equation}
    H_\drive = \hbar\left(\sum_{i=1}^3 g_i \cos(\phi_i)\cos(\theta_i) \right)\hat N
\end{equation}
where $\theta_i(t)=2\pi f_i t$ is the phase of the drive and $\phi_i(t)=\omega_i t$ is the phase of time-modulation of the amplitude of the drive.
The term in parenthesis is proportional to the amplitude of the propagating wave on the line
\begin{align}
    \mathcal{A}(t) &= \sum_{i=1}^3 2\mathcal{A}_i\cos(\phi_i(t))\cos(\theta_i(t)) \\
    &= \sum_{i=1}^3 \mathcal{A}_i\left( \cos(\theta_i^+(t)) + \cos(\theta_i^-(t)) \right)
\end{align}
with~$\theta_i^\pm=\theta_i\pm\phi_i$. Thus, the transmission line contains six modes at frequencies~$f_i^\pm=f_i\pm\omega_i/2\pi$, for $i=1,2,3$ (see Fig.~\ref{fig:exp}).
As explained in Sec.~\ref{sec:exp}, we model the propagating mode at frequency~$f_i^\pm$ as a classical mode of energy~$hfn_i^\pm$, where the phase~$\theta_i^\pm$ of each mode is conjugated to~$\hbar n_i^\pm$,
such that the net photon flux is given by the difference between the outgoing and incoming signals at this frequency~$h f_i^\pm \dot n_i^\pm = S_\text{out}[f_i^\pm] - S_\text{in}[f_i^\pm]$.
The topological pumping describes the dynamics of the variables conjugated to~$\phi_i$.
The change of variables
\begin{align}
    \theta_i &= \frac{1}{2}\left( \theta_i^+ + \theta_i^- \right) \\
    \phi_i &= \frac{1}{2}\left( \theta_i^+ - \theta_i^- \right)\\
    m_i &=  n_i^+ + n_i^-  \\
    n_i &=  n_i^+ - n_i^- 
\end{align}
is a canonical change of variables, which means that it preserves the Poisson brackets, so the variable~$\hbar n_i = \hbar(n_i^+ - n_i^-)$ is conjugated to the phase $\phi_i$.
The topological pumping relates the rates~$\dot n_i$, so we want to measure
\begin{align}
    \dot{n}_i
    &=\frac{S_\text{out}[f_i^+]-S_\text{in}[f_i^+]}{h f_i^+}-\frac{S_\text{out}[f_i^-]-S_\text{in}[f_i^-]}{h f_i^-} \\
    &\simeq \frac{\Delta S_i}{h f_i}
\end{align}
with $\Delta S_i=S_\text{out}[f_i^+]-S_\text{out}[f_i^-]$ if we consider~$f_i^\pm\simeq f_i$ and $S_\text{in}[f_i^+]=S_\text{in}[f_i^-]$.

\section{Chern insulator on the Lieb lattice}
\label{annexe:LiebLatticeAnalysis}

The three-band insulator on the Lieb lattice satisfies inversion symmetry. We write the inversion operator as $P(\boldsymbol{k})$=diag($e^{ik_{x}}$,1,$e^{ik_{y}}$). It leads to four inversion-invariant momenta in the 2D Brillouin zone: $\boldsymbol{\Gamma}=(0,0)$, $\boldsymbol{X}=(\pi,0)$, $\boldsymbol{Y}=(0,\pi)$, and $\boldsymbol{M}=(\pi,\pi)$. At these high-symmetry points, the Bloch Hamiltonian commutes with the inversion operator. Thus, the parities $p_\nu$ -- eigenvalues of the inversion operator -- are good quantum numbers to label each energy band $\nu$ at the inversion-invariant momenta. We sort the energy bands as $E_0<E_1<E_2$ and introduce the triplet $\boldsymbol{\pi}=\left(\pi_0,\pi_1,\pi_2\right)$, where $\pi_\nu$ is the band parity product
\begin{align}
    \pi_\nu = p_\nu(\boldsymbol{\Gamma})p_\nu(\boldsymbol{X})p_\nu(\boldsymbol{Y})p_\nu(\boldsymbol{M}) = \pm1 ~.
\end{align}
We now determine the different configurations of parity products allowed for the three bands emulated by the fluxonium.
\\

At momentum $\boldsymbol{\Gamma}$, the parity operator reduces to the identity matrix. All energy bands have the same parity, regardless of the Hamiltonian parameters. This leads to the parity triplet $\boldsymbol{p}(\boldsymbol{\Gamma})=(p_0(\boldsymbol{\Gamma}),p_1(\boldsymbol{\Gamma}),p_2(\boldsymbol{\Gamma}))=(+,+,+)$.
\\

\begin{figure*}[t]
\centering
\includegraphics[trim = 0mm 0mm 0mm 0mm, clip, width=12cm]{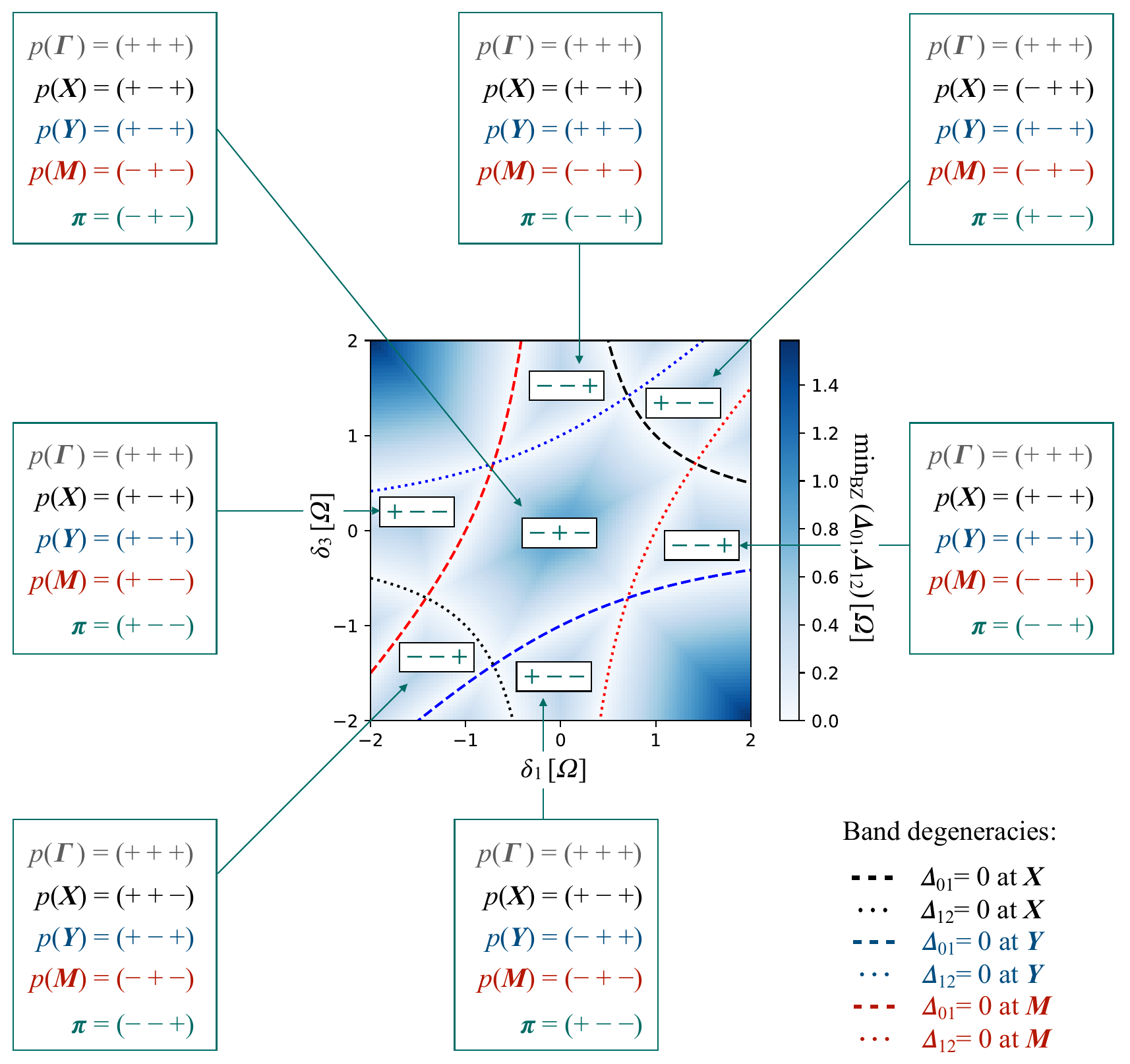}
\caption{\small \textbf{Parity product of the bands}: The central map represents the minimum value of the lower gap ($\Delta_{01}$) or upper gap ($\Delta_{12}$) over the Brillouin zone, as a function of the detunings $\delta_1$ and $\delta_3$. It is obtained from numerical diagonalization of the Bloch Hamiltonian (\ref{HLieb}). Energy is in units of $\Omega_{1}=\Omega_{2}=\Omega_{3}=\Omega$. The white areas correspond to values of the detuning for which a band gap closes.
They are well described by the dashed (dotted) lines obtained analytically from the closing conditions of gap $\Delta_{01}$ ($\Delta_{12}$) at the inversion-invariant momenta $\boldsymbol{\Gamma}$, $\boldsymbol{X}$, $\boldsymbol{Y}$, and $\boldsymbol{M}$ in the Brillouin zone. These degeneracy lines mark the transitions between topologically nonequivalent band insulators, where the parity product $\pi_\nu$ of certain bands changes signs. The inversion eigenvalues labeling the bands at $\boldsymbol{\Gamma}$, $\boldsymbol{X}$, $\boldsymbol{Y}$, and $\boldsymbol{M}$ are shown in the green boxed insets, as a parity triplet $\boldsymbol{p}=(p_0,p_1,p_2)$ associated with the band energies $E_0<E_1<E_2$. It is only shown for insulating regions that exhibit negative band parity products. In such regions, the band structures do not support any band representation and cannot be adiabatically connected to an atomic limit. The fluxonium qutrit then simulates nontrivial Chern insulators.}
\label{NonBandRepresentations}
\end{figure*}

At momentum $\boldsymbol{X}$, the parity operator and the Bloch Hamiltonian read $P(\boldsymbol{X})=\text{diag}(-1,1,1)$ and
\begin{align}
    H(\boldsymbol{X}) = 
    \left(
    \begin{array}{ccc}
    0 & 0 & 0 \\
    0 & \delta_{1} & \Omega_{2} \\
    0 & \Omega_{2} & \delta_{3} \\
    \end{array}
    \right) . \notag
\end{align}
The eigenspace of parity $-1$ is associated with the energy level $E_{*}(\boldsymbol{X})=0$. This is fixed regardless of the Hamiltonian parameters. In contrast, the eigenspace of parity $+1$ refers to the energy levels
\begin{align}
    E_{\pm}(\boldsymbol{X})=\frac{\delta_{1}+\delta_{3}}{2} \pm \frac{1}{2}\sqrt{4\Omega_{2}^{2}+(\delta_{1}-\delta_{3})^{2}} ~,
\end{align}
which depends on the fluxonium drives. This allows three different configurations of parities:
\begin{itemize}
\item $E_{*}(\boldsymbol{X})<E_{-}(\boldsymbol{X})<E_{+}(\boldsymbol{X})$ with parities $\boldsymbol{p}(\boldsymbol{X})=(-,+,+)$. It occurs when $\delta_{1}>0$, $\delta_{3}>0$, and $\delta_{3}>\Omega_{2}^{2}/\delta_{1}$.
\item $E_{-}(\boldsymbol{X})<E_{+}(\boldsymbol{X})<E_{*}(\boldsymbol{X})$ with parities $\boldsymbol{p}(\boldsymbol{X})=(+,+,-)$. It occurs when $\delta_{1}<0$, $\delta_{3}<0$, and $\delta_{3}<\Omega_{2}^{2}/\delta_{1}$.
\item $E_{-}(\boldsymbol{X})<E_{*}(\boldsymbol{X})<E_{+}(\boldsymbol{X})$ with parities $\boldsymbol{p}(\boldsymbol{X})=(+,-,+)$ otherwise.
\end{itemize}

At momentum $\boldsymbol{Y}$, the parity operator and the Bloch Hamiltonian read $P(\boldsymbol{Y})=\textrm{diag}(1,1,-1)$ and
\begin{align}
    H(\boldsymbol{Y}) = 
    \left(
    \begin{array}{ccc}
    0 & \Omega_{1} & 0 \\
    \Omega_{1} & \delta_{1} & 0 \\
    0 & 0 & \delta_{3} \\
    \end{array}
    \right) . \notag
\end{align}
The eigenspace of parity $-1$ is associated with the energy level $E_{*}(\boldsymbol{Y})=\delta_{3}$. The eigenspace of parity $+1$ refers to the energy levels
\begin{align}
E_{\pm}(\boldsymbol{Y})=\frac{\delta_{1}}{2} \pm \frac{1}{2}\sqrt{4\Omega_{1}^{2}+\delta_{1}^{2}} ~.
\end{align}
This allows three different configurations of parities:
\begin{itemize}
\item $E_{*}(\boldsymbol{Y})<E_{-}(\boldsymbol{Y})<E_{+}(\boldsymbol{Y})$ with parities $\boldsymbol{p}(\boldsymbol{Y})=(-,+,+)$. It occurs when $2\delta_{3}<\delta_{1}-\sqrt{4\Omega_{1}^{2}+\delta_{1}^{2}}$.
\item $E_{-}(\boldsymbol{Y})<E_{+}(\boldsymbol{Y})<E_{*}(\boldsymbol{Y})$ with parities $\boldsymbol{p}(\boldsymbol{Y})=(+,+,-)$. It occurs when $2\delta_{3}>\delta_{1}+\sqrt{4\Omega_{1}^{2}+\delta_{1}^{2}}$.
\item $E_{-}(\boldsymbol{Y})<E_{*}(\boldsymbol{Y})<E_{+}(\boldsymbol{Y})$ with parities $\boldsymbol{p}(\boldsymbol{Y})=(+,-,+)$ otherwise.
\end{itemize}

At momentum $\boldsymbol{M}$, the parity operator and the Hamiltonian read $P(\boldsymbol{M})=\textrm{diag}(-1,1,-1)$ and
\begin{align}
    H(\boldsymbol{M}) = 
    \left(
    \begin{array}{ccc}
    0 & 0 & -i\Omega_{3} \\
    0 & \delta_{1} & 0 \\
    i\Omega_{3} & 0 & \delta_{3} \\
    \end{array}
    \right) . \notag
\end{align}
The eigenspace of parity $+1$ is associated with the energy level $E_{*}(\boldsymbol{M})=\delta_{1}$. The eigenspace of parity $-1$ refers to the energy levels
\begin{align}
E_{\pm}(\boldsymbol{M})=\frac{\delta_{3}}{2} \pm \frac{1}{2}\sqrt{4\Omega_{3}^{2}+\delta_{3}^{2}} ~.
\end{align}

This allows three different energy configurations:
\begin{itemize}
\item $E_{*}(\boldsymbol{M})<E_{-}(\boldsymbol{M})<E_{+}(\boldsymbol{M})$ with parities $\boldsymbol{p}(\boldsymbol{M})=(+,-,-)$. It occurs when $\delta_{1}<0$ and $\delta_{3}>\delta_{1}-\Omega_{3}^{2}/\delta_{1}$.
\item $E_{-}(\boldsymbol{M})<E_{+}(\boldsymbol{M})<E_{*}(\boldsymbol{M})$ with parities $\boldsymbol{p}(\boldsymbol{M})=(-,-,+)$. It occurs when $\delta_{1}>0$ and $\delta_{3}<\delta_{1}-\Omega_{3}^{2}/\delta_{1}$.
\item $E_{-}(\boldsymbol{M})<E_{*}(\boldsymbol{M})<E_{+}(\boldsymbol{M})$ with parities $\boldsymbol{p}(\boldsymbol{M})=(-,+,-)$ otherwise.
\end{itemize}

This shows that any change of band parity requires the band gap to close at the inversion-invariant momenta. All the band-parity configurations determined above, as well as their parity products, are summarized in the insets of Fig.~\ref{NonBandRepresentations} as a function of $\delta_1$ and $\delta_3$.

\section{Dynamics in the rotating frame}\label{annexe:dynamicsRotatingFrame}
The equations of dynamics of the classical variables~$n_i$ conjugated to the phases~$\phi_i$ are first written in the laboratory frame
\begin{equation}
    \dot n_i = -\frac{1}{\hbar} \bra{\Psi_\labo(t)} \frac{\partial H_\labo}{\partial \phi_i} \ket{\Psi_\labo(t)},
\end{equation}
where the dynamics of state~$\ket{\Psi_\labo(t)}$ of the qutrit in the laboratory frame is governed by the Hamiltonian~$H_\labo(\{\phi_j, \theta_j\})$, introduce before~\eqref{eq:Hlabo}.
In the rotating frame, the dynamics of~$\ket{\Psi_\rot(t)}=U^\dagger(t)\ket{\Psi_\labo(t)}$ is governed by
$H_\rot=U^\dagger H_\labo U - i\hbar U^\dagger\frac{\dd U}{\dd t}$
which satisfies~$\frac{\partial H_\rot}{\partial\phi_i}=U^\dagger\frac{\partial H_\labo}{\partial \phi_i}U$ since the unitary transformation~$U(t)$ does not depend on the phase~$\phi_i$. Thus, the equation of the dynamics of~$n_i$ has the same form in the rotating frame
\begin{equation}
    \dot n_i = -\frac{1}{\hbar} \bra{\Psi_\rot(t)} \frac{\partial H_\rot}{\partial \phi_i} \ket{\Psi_\rot(t)}
\end{equation}
where in the rotating wave approximation, we consider $H_\rot(\{\phi_j,\theta_j\},t)\simeq \tilde H(\phi_1, \phi_2, \phi_3)$, with the 3 phases Hamiltonian given by~\eqref{eq:Hphi1phi2phi3}.
We consider the following canonical change of variable
 \begin{subequations}
 \begin{align}
 n_I     &= n_1 + n_3  &\quad ; \quad \phi_{I}   &= \phi_1 \\
 n_{II}  &= n_2 - n_3  &\quad ; \quad \phi_{II}  &= \phi_2\\
 n_{III} &= -n_3       &\quad ; \quad \phi_{III} &= \phi_1 - \phi_2 - \phi_3
 \end{align}  
 \end{subequations}
satisfying~$\{\phi_A, n_B\}=1/\hbar$, $A,B=I,II,III$. Since these new variables are conjugated, the equations of motion are
\begin{equation}
    \dot{n}_A  = -\frac{1}{\hbar}\frac{\partial E_\nu}{\partial \phi_A} 
    +\sum_{B\neq A} \dot\phi_B\Berry^{(\nu)}_{\phi_A \phi_B}
\end{equation}
with~$E_\nu$ the energy and~$\Berry^{(\nu)}$ the Berry curvature of the band~$\nu$ of
$H(\phi_I, \phi_{II},\phi_{III})=\tilde H(\phi_I, \phi_{II}, \phi_I-\phi_{II}-\phi_{III})$.
The frequencies of the phases~$\phi_1, \phi_2, \phi_3$ are chosen such that~$\omega_3=\omega_1-\omega_2$, thus~$\dot\phi_{III}=0$ and we keep~$\phi_{III}=0$ at all time with the initial condition. Thus, the equations of motion reduce to~\eqref{eq:nI_nII_dynamics}.

\section{Temporal fluctuations}\label{annexe:TempFluctuations}
During the adiabatic evolution, the time derivative of the energy of a mode can be decomposed in a sum of three terms
\begin{equation}
    \hbar\omega_1(\dot n_1+\dot n_3) = -\omega_1\frac{\partial E_\nu}{\partial\phi_1}
    + \hbar\omega_1 \omega_2 (F_{\phi_1 \phi_2}^{(\nu)} - \frac{\Chern^{(\nu)}}{2\pi})
    + \hbar\frac{\omega_1\omega_2}{2\pi}\Chern^{(\nu)}.
\end{equation}
The first term is the variation of the energy~$E_\nu$ of the band~$\nu$ of the qutrit, corresponding to an energy exchange between the qutrit and the mode. 
The second term correspond to the fluctuation of the Berry curvature~$\Berry^{(\nu)}_{\phi_1 \phi_2}$ around its topologically quantized average value~$\frac{\Chern^{(\nu)}}{2\pi}$, with~$\Chern^{(\nu)}$ the Chern number.
This corresponds to the fluctuation of the geometrical transfer of energy between the two modes.
The last term is the topological power rate, the only non-zero term in time-average.
The two first terms are responsible for the time fluctuation of the energy of the mode.

\begin{figure}[t]
\includegraphics[trim = 0mm 0mm 0mm 0mm, clip, width=8.5cm]{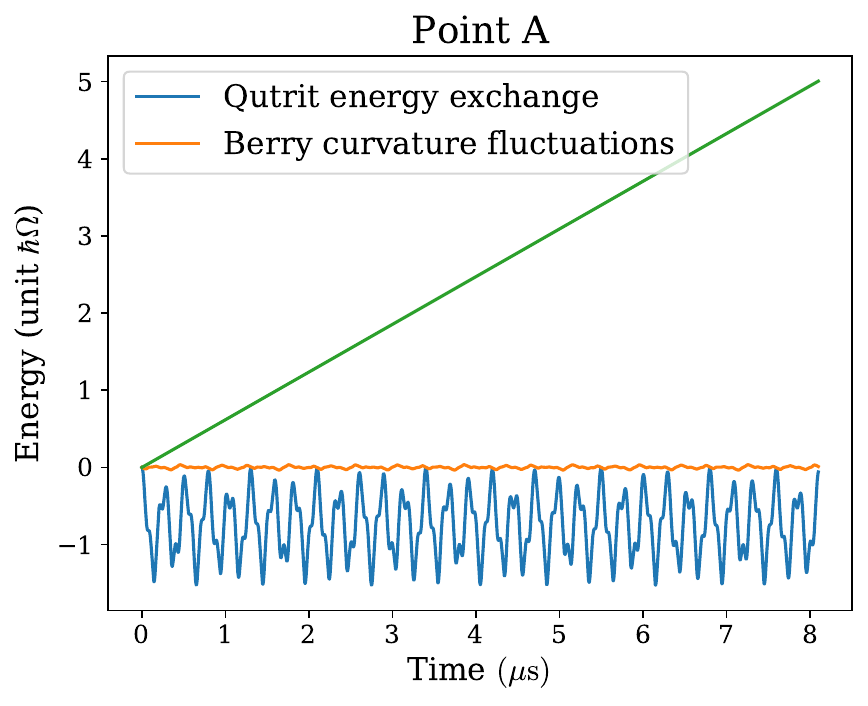}
\caption{Different terms in the variation of the energy~$\hbar\omega_1 (n_1+n_3)$, in the case of resonance $\delta_1=\delta_3=0$, point~A in Fig.~\ref{fig:TopTransition}(a). In blue: Time-integration of the term of variation of the energy of the qutrit $-\omega_1\frac{\partial E_\nu}{\partial\phi_1}$. In orange: Time-integration of the term of fluctuation of the geometrical coupling~$\hbar\omega_1 \omega_2 (F_{\phi_1 \phi_2}^{(\nu)} - \frac{\Chern^{(\nu)}}{2\pi})$. In green: Topological energy transfer at constant rate~$\hbar\frac{\omega_1\omega_2}{2\pi}\Chern^{(\nu)}$.
The fluctuation of the energy of the qutrit is the predominant source of temporal fluctuation of the energy.}
\label{fig:TemporalFluctuations}
\end{figure}

In Fig.~\ref{fig:TemporalFluctuations} is represented the time-integration of each term in the case of resonance~$\delta_1=\delta_3=0$, point~A in Fig.~\ref{fig:TopTransition}(a). We see that the temporal fluctuation of the energy is mainly due to the energy exchange between the qutrit and the mode, and the fluctuation of the Berry curvature is much lower. 
This is the case for every value of parameters in the region of interest of the phase diagram.

\end{document}